\documentclass[12pt,amsmath,amssymb,showkeys,prb]{revtex4}
\usepackage[T1]{fontenc}
\usepackage{xcolor}
\usepackage{setspace}
\usepackage{graphicx,caption}
\newcommand{\citen}[1]{%
  \begingroup
    \romannumeral-`\x 
    \setcitestyle{numbers}%
    \cite{#1}%
  \endgroup
}

\usepackage{amssymb}
\usepackage{amsmath}
\usepackage{bbold}

\usepackage{mathrsfs}
\usepackage{calligra}
\usepackage{bm}

\usepackage{xcolor}
\usepackage{hyperref}

\newcommand{\e}{\mathrm e}
\renewcommand{\d}{\mathrm d}
\renewcommand{\i}{\mathrm i}

\newcommand{\rhop}{ \rho_{\rm p} }

\newcommand{\rr}{ {\bf r} }

\newcommand{\kk}{ {\bf k} }
\newcommand{\RR}{ {\bf R} }

\newcommand{\D}{ {\cal D} }
\newcommand{\lB }{ l_{\rm B} }

\newcommand{\np}{n_{\rm p}} 
\newcommand{\nw}{n_{\rm w}} 

\begin{document}

$\null$
\hfill {November 8, 2022}
\vskip 0.3in

\begin{center}
{\Large\bf Analytical Formulation and Field-Theoretic}\\ 

\vskip 0.3cm

{\Large\bf Simulation of Sequence-Specific Phase Separation of}\\ 

\vskip 0.3cm

{\Large\bf Proteinlike Heteropolymers with Short- and}\\

\vskip 0.3cm

{\Large\bf Long-Spatial-Range Interactions}\\

\vskip .5in
{\bf Jonas W{\footnotesize{\bf{ESS\'EN}}}},$^{}$
{\bf Suman D{\footnotesize{\bf{AS}}}},$^{}$
{\bf Tanmoy P{\footnotesize{\bf{AL}}}},$^{}$
 and
{\bf Hue Sun C{\footnotesize{\bf{HAN}}}}$^{*}$

$\null$

Department of Biochemistry,
University of Toronto, Toronto, Ontario M5S 1A8, Canada\\

\vskip 1.3cm

%

\end{center}

\vskip 1.3cm

\noindent
\\

\noindent
$*$Corresponding author\\
{\phantom{$^\dagger$}}
E-mail: huesun.chan@utoronto.ca;
Tel: (416)978-2697; Fax: (416)978-8548\\
{\phantom{$^\dagger$}}
URL:
\href{http://biochemistry.utoronto.ca/person/hue-sun-chan/}
{\tt http://biochemistry.utoronto.ca/person/hue-sun-chan/}\\
Mailing address:\\
{\phantom{$^\dagger$}}
Department of Biochemistry, University of Toronto,
Medical Sciences Building -- 5th Fl.,\\
{\phantom{$^\dagger$}}
1 King's College Circle, Toronto, Ontario M5S 1A8, Canada.\\

\vskip 2cm

\centerline{{\tt Accepted for publication in }{\it J Phys Chem B} {\tt as an
article in the upcoming}}
\centerline{{\tt "Jos\'e Onuchic Festschrift" Special Issue}}

\vfill\eject

\noindent
{\large\bf Abstract}\\

\noindent
A theory for sequence dependent liquid-liquid phase separation (LLPS) of
intrinsically disordered proteins (IDPs) in the study of biomolecular
condensates is formulated by extending the random phase approximation  (RPA)
and field-theoretic simulation (FTS) of heteropolymers with spatially
long-range Coulomb interactions to include the fundamental effects of
short-range, hydrophobic-like interactions between amino acid residues. To this
end, short-range effects are modeled by Yukawa interactions between multiple
nonelectrostatic charges derived from an eigenvalue
decomposition of pairwise residue-residue contact energies. Chain excluded
volume is afforded by incompressibility constraints. A mean-field approximation
leads to an effective Flory $\chi$ parameter, which, in conjunction with RPA,
accounts for the contact-interaction effects of amino acid composition and the
sequence-pattern effects of long-range electrostatics in IDP LLPS, whereas FTS
based on the formulation provides full sequence dependence for both short- and
long-range interactions. This general approach is illustrated here by
applications to variants of a natural IDP in the context of several different
amino-acid interaction schemes as well as a set of different model 
hydrophobic-polar sequences sharing the same composition. Effectiveness 
of the methodology is verified by coarse-grained explicit-chain molecular 
dynamics simulations.

\vfill\eject

\noindent
{\large\bf INTRODUCTION}\\

Tremendous recent advances have made it abundantly clear that biomolecular 
condensates serve many important biological 
functions.\cite{rosen2017,cliff2017,rosen2021}
The architecture of functional intraorganismic biomolecular condensates 
can be highly complex. Some contain hundreds of different kinds of 
proteins---including folded protein domains, intrinsically disordered
proteins (IDPs) and intrinsically disordered protein regions (IDRs)---and 
various nucleic acids participating in an intricate network of interactions,
as exemplified by the $\sim 300$ protein species identified from proteomic 
analyses of yeast and mammalian stress granules using mass 
spectrometry.\cite{parker2016}
While the existence of intracellular liquid-like compartments has been 
suggested more than 120 years ago from observing that the protoplasm of
starfish and sea urchin eggs resembles ``a mixture of liquids'' with 
suspended droplets of different chemical nature,\cite{wilson1899}
the intense modern interest in biomolecular condensates was sparked by
the recognition that the assembly of these condensates
bears close resemblance to the physical process of liquid-liquid phase 
separation (LLPS)---as noted in a seminal study of germline P granules 
a little more than a decade ago.\cite{brangwynne2009} 
The LLPS perspective has since contributed greatly to the discovery and
understanding of a large variety of biomolecular condensates. Progress is often
achieved through experimental characterizations of the phase behaviors as well 
as the biochemical and materials properties of constructs, composing of 
only a few species of protein and nucleic acid, to serve as models of the 
biomolecular condensates of interest.\cite{Rosen12,McKnight12,Nott15,tanja2015,cliff2015,parker2015,Michnick2016,babu2018,Monika2018Rev,Roland2019,shorter2019}
One novel observation from these investigations is that multivalent, relatively 
weak interactions of IDPs and IDRs---typically weaker than binding of
folded domains---often play a major role in the formation 
of biomolecular condensates.\cite{McKnight12,Nott15,tanja2015,parker2015}
These interactions among largely disordered conformations of IDPs and IDRs
are transient and presumably nonstoichiometric,\cite{Fawzi2015,jacob2017}
though in some instances they entail sampling disordered chain 
configurations with labile but nonetheless specific fibril-like 
local structures~\cite{McKnight12,McKnight2018}. Despite
the stochastic nature of these IDP/IDR interactions, they are capable of
contributing to a ``fuzzy'', sequence-dependent mechanism of molecular 
recognition.\cite{feric2016,njp2017}

Confronted by the overwhelming complexity of functional intraorganismic 
condensates, it goes without saying that few-component LLPS under 
thermodynamic equilibrium is a rudimentary---yet important---model 
for gaining insights into biomolecular condensates in living organisms.
For instance, nonequilibrium behaviors of {\it in vivo} condensates are
more appropriately viewed as those of active liquids;\cite{julicher2018,lee2022}
and formulations beyond simple LLPS theory are needed to describe
the size, viscoelastic, and other materials properties of
biomolecular condensates.\cite{dufresne2020,hxzhou2021,espinosa2021}
In addition to physicochemical processes akin to LLPS, gelation and 
percolation\cite{RohitRosen2017,biochemrev}
as well as structure-specific stoichiometric 
interactions\cite{Tjian2019,musacchio2022} are expected to also contribute 
significantly to the formation of biomolecular condensates. It is
likely that these processes are coupled\cite{Rohit2022} as, at least
in the example of a model condensate for postsynaptic densities, 
stoichiometric interactions alone do not account for the experimentally 
observed phase properties.\cite{YHLin2022}
Nonetheless, despite these complexities, since LLPS is integral to---though
not exclusively responsible for---biomolecular condensate formation,
experimental\cite{brangwynne2009,Rosen12,McKnight12,Nott15,tanja2015,cliff2015,parker2015,Michnick2016,babu2018,Monika2018Rev,Roland2019,shorter2019}
and theoretical/computational\cite{njp2017,CellBiol,NatPhys,linPRL,linJML,lin2017,dignon18,suman1,jeetainPNAS,suman2,stefan2019,joanElife,joanJPCL}
studies of equilibrium LLPS is a valuable tool for gaining physical, 
chemical, and biological insights. In this regard, they play an essential
role similar to other simple model systems in many branches of 
science---including, as a befitting example for this Special Issue, 
a seminal lattice model study by Onuchic and coworkers\cite{leopold} 
that inspired the funnel landscape
picture of protein folding,\cite{wolynes,dillchan97}
with the obvious proviso that many aspects of the recent simple model
LLPS  systems' quantitative relationship with the biological functions of 
intraorganismic biomolecular condensates remain to be delineated.
Indeed, the LLPS perspective has already led to several notable
conceptual advances,
including a likely role of the general physical principle of phase separation 
in biomolecular homeostasis,\cite{njp2017,biochemrev,Zechner2020,safran2021}
novel clues to neurological effects
of hydrostatic pressure,\cite{roland2020,roland2022}
and a likely link between neurological disease-causing mutations 
and bioinformatics-inferred LLPS propensities,\cite{BrianJulie2020}
to name a few.

Depending on the phenomena and scientific questions of interest,
theoretical/computational modeling of biomolecular LLPS may 
endeavor to capture different levels of 
structural and energetic detail.\cite{biochemrev} 
These modeling approaches include---but certainly not limited to---mean-field 
Flory-Huggins (FH)\cite{FH-ref} and Overbeek-Voorn (OV)\cite{OV-ref}
theories of polymer solutions,
random phase approximation (RPA) theories to address effects of
sequence charge patterns on LLPS of polyampholytes as model
IDPs/IDRs,\cite{linPRL,linJML,delacruz2003}
field-theoretic simulation (FTS)\cite{Fredrickson2006}
to provide an improved account of charge and matter density fluctuations
beyond RPA for polyampholyte LLPS\cite{joanElife,joanJPCL,joanPNAS,joanJCP}---though FTS is still limited by finite-size effects and in its treatment
of excluded volume\cite{irback2020,Pal2021,irback2021,irback2022}---as well 
as explicit-chain simulations of 
lattice,\cite{feric2016,suman1,stefan2019,lassi2019}
continuum coarse-grained,\cite{dignon18,jeetainPNAS,suman2,SumanPNAS,panag2017,jeetain-rev2021}
and atomistic\cite{regis2017,JeetainAtom}
models of IDP/IDR LLPS.
Transfer matrix, restricted primitive model simulations,
and other techniques have been applied to study complex coacervation of 
polyampholytes and polyelectrolytes;\cite{transferM2017,singperry2017,sing2020}
and LLPS of single or multiple biomolecular species, especially those 
involving folded proteins and folded domains, have also been modeled by 
patchy particles\cite{vlachy2016,hxzhou2018,roxana2020}
using simulations\cite{hxzhou2018,roxana2020}
as well as analytical formalisms based on Wertheim’s thermodynamic 
perturbation theory.\cite{vlachy2016,wertheim1986}
As expected, the required computation
increases with structural and energetic
details that a model seeks to capture. In this context, it is noteworthy
that even basic FH theory, which requires minimal numerical 
effort\cite{mimb2022} yet has recently been made even more tractable
by a novel self-consistent solution,\cite{analyticalFH}
can be very useful in advancing knowledge about biomolecular LLPS.
This is exemplified by the applications of FH to delineate
scenarios of tie-line patterns and their ramifications for
homeostasis,\cite{njp2017,biochemrev,safran2021}
to ascertain the extent of void-volume contributions to the 
hydrostatic pressure dependence of LLPS,\cite{roland2020} 
and to rationalize experimental data on the impact of
aromatic valence on IDP LLPS.\cite{TanjaScience2020}

A fundamental limitation of mean-field FH and OV theories is that
they consider the composition of heteropolymer sequences 
without accounting for the full effects of the sequential arrangements
of monomers (e.g., amino acid residues for IDPs/IDRs) along heteropolymer 
chains. While explicit-chain simulations of LLPS readily embody 
sequence-pattern effects,\cite{dignon18,suman1,jeetainPNAS,suman2,stefan2019,regis2017,SumanPNAS,JeetainAtom,Alan2020,koby2020,Davit2020,koby2022,Jeetain-domains} 
theories that take into account
sequence patterns but are computationally less intensive have proven 
useful as complementary approaches, especially for 
screening large number of sequences. To date, these 
sequence-specific 
theories, which include RPA, FTS, variational approaches,\cite{kings2015}
and RPA augmented by Kuhn-length renormalization (rG-RPA)\cite{rG-RPA-ref} 
for a more physical account of charge density and 
its fluctuations,\cite{dePablo2022}
have focused only on spatially long-range Coulomb interactions entailed
by the heteropolymeric sequence patterns of electric 
charges\cite{kings2015,rohit2013,Kings2022}
and their effects
on LLPS,\cite{linPRL,linJML,lin2017,joanElife,joanJPCL,joanPNAS,joanJCP,Pal2021,mimb2022,rG-RPA-ref,wessen2021,wessen2022}
single-chain polyampholyte/IDP conformational 
properties,\cite{lin2017,dePablo2022,Kings2022,kings2017,firman2018,huihui2018,kings2020} and the interaction of a pair of polyampholyte chains.\cite{Alan2020}

These theoretical treatments of spatially long-range electrostatic 
interactions have led to many important physical insights; but
these theories by themselves do not address the full sequence-dependent 
effects of spatially short-range, contact-like hydrophobic\cite{regis2017} and 
$\pi$-related\cite{SongKAW2013,robert} interactions, which
are integral parts of the physical driving forces for biomolecular
LLPS\cite{dignon18,SumanPNAS,regis2017,robert,moleculargrammar,kitahara2021,kameda2022}
as well as IDP conformational 
properties\cite{jeetainPNAS,zhengHP,song21} and interactions.\cite{SongKAW2013}
In lieu of an account of sequence-pattern dependence,
short-range interactions are sometimes treated by augmenting a 
sequence-dependent theory for electrostatic interactions such as RPA by a 
composition-dependent mean-field FH account of spatially short-range 
interactions.\cite{linPRL,linJML}
One source of difficulty in extending RPA to include spatially short-range
attractive (favorable) interactions is that they are prone to produce
mathematical singularities in the RPA free energy, signalling the instability
of the homogeneous phase about which the RPA expansion is performed. This
observation led us to consider the possibility of incorporating these
interactions in FTS because FTS does not require the approximations in RPA
theory and should, therefore, properly handle the singularities.
We put forth such a theory here.
To account for nonelectrostatic, spatially short-range interactions
in our FTS formulation, additional fields are introduced, wherein 
different amino acid residues are assigned different nonelectrostatic charges
associated with the newly introduced fields. 
The nonelectrostatic charges are determined as components of eigenvectors
using eigenvalue decomposition of the pairwise interaction 
energies\cite{HaoLi1997,chan99,CieplakJCP2001,krestenBJ2008} 
from either the values for the 20 amino 
acids on a hydrophobicity (hydropathy) 
scale\cite{dignon18,Rossky,Urry,LZ2020,Urry-Mittal,FB,krestenPNAS2021}
or the $20\times 21/2=210$ pairwise contact energies
between the amino acids (the interaction matrix) derived from
statistical analyses of the Protein Data Bank (PDB) and/or from physical 
considerations.\cite{dignon18,MJ85,MJ96,KH,Mpipi,EDLevy2022}
This approach is computationally tractable because matrices for physical 
pairwise interactions among amino acid residues tend to possess
a dominant eigenvector.\cite{HaoLi1997,chan99,godzik1995,CieplakJCP2001,chanMIT}
Thus, only a small number of types of nonelectric charge ($\ll 20$) 
are necessary to provide a mathematically good approximation to
the $210$ pairwise interactions.
In view of the long-standing recognition that there are significant 
discrepancies among hydrophobicity scales because of differences
in the experimental or computational techniques used for their
construction,\cite{pakarplus1997,devido1998,Pappu2021}
the size- and length-scale dependence of hydrophobic 
interactions\cite{chandler1999,scheraga2007,scheraga2008}
and their deviations from pairwise 
additivity\cite{chan2011,shimizu01,shimizu02}
(as manifested, e.g., by the fact that pairwise interactions 
alone are insufficient to account for protein folding 
cooperativity\cite{chan2011,Chan2004,chan98}), we apply our new 
formulation to several different hydrophobicity 
scales\cite{dignon18,Urry-Mittal,FB}
and pairwise interaction schemes\cite{dignon18,Mpipi}
to compare their predicted LLPS behaviors for the wildtype and variant 
IDRs of the DEAD-box RNA helicase Ddx4\cite{jacob2017,Nott15}
and two-letter (hydrophobic-polar) model sequences.\cite{Statt2020,panagio2021}
The theoretical predictions are compared against experimental trend and
corresponding predictions from coarse-grained explicit-chain 
molecular dynamics simulations.
These results, together with their underpinning theoretical development and 
their ramifications for future efforts, are provided in detail below.
\\


$\null$

\noindent
{\large\bf MODELS AND METHODS}\\

We begin by presenting a general analytical formulation, amenable
to RPA and FTS, for a model system containing multiple chains of 
heteropolymers with specific sequences of residues (monomers) together 
with small ions and solvent molecules, wherein the constituents may interact 
via spatially short-range, contact-like interactions as well as 
spatially long-range Coulomb interactions.\\


{\bf Spectral (Eigenvalue) Decomposition of Residue-Residue Energy Matrices
and Definition of Nonelectric Charges.} 
Assuming that the spatially short-range interactions among the 
twenty types of amino acid residues may be described approximately
by a $20\times 20$ matrix $\varepsilon_{r,r'}$ of contact energies 
for the pairwise interactions between residue types $r$ and $r'$, 
this matrix can be written in the diagonalized form
\begin{equation} \label{eq:eps_spectral_decomposition}
\varepsilon_{r,r'} = \sum_{a=1}^{20} \lambda_a q^{(a)}{}_r q^{(a)}{}_{r'} 
\end{equation}
because $\varepsilon_{r,r'}$ is symmetric 
($\varepsilon_{r,r'}=\varepsilon_{r',r}$), with
$\lbrace \lambda_a\rbrace_{a=1}^{20}$ being the eigenvalues of 
$\varepsilon_{r,r'}$, and $q^{(a)}{}_{r}$ constituting a set of twenty 
orthogonal, unit-normalized eigenvectors labeled by $a$. 
We may then view the eigenvector component $q^{(a)}{}_{r}$ as the
value of a mathematically defined type-$a$ nonelectric charge for residue 
type $r$. Without loss 
of generality, we label the eigenvalues (and thus the nonelectric charge type) 
such that the eigenvalues are ordered according to descending magnitude,
i.e., $|\lambda_a| \geq |\lambda_{a+1} |$. 

The spectral form in Eq.~\eqref{eq:eps_spectral_decomposition} implies that we
can decompose the full $\varepsilon_{r,r'}$ into a set of interactions, with
overall strengths $\lambda_a$ and relative residue-dependent strengths given
by product of nonelectric charges $q^{(a)}{}_{r} q^{(a)}{}_{r'}$. 
Indeed, this approach has long been used to construct hydrophobicity scales 
using the eigenvalue decomposition of statistical contact potentials such as 
the classic Miyazawa-Jernigan (MJ) matrix.\cite{HaoLi1997,chan99,MJ85,MJ96}

Although all twenty eigenvalues are needed 
in Eq.~\eqref{eq:eps_spectral_decomposition} to recover the 
full $\varepsilon_{r,r'}$ exactly, one can often achieve a very 
good numerical agreement by including only the first few most 
dominant $\lambda_a$, i.e.~by truncating the sum at some $\tilde{a}< 20$,
\begin{equation} \label{eq:truncated}
\sum_{a=1}^{20} \rightarrow \sum_{a=1}^{\tilde{a}} ,
\end{equation}
because of the pattern of hydrophobic-like interactions among amino
acid residues.\cite{HaoLi1997,chan99,godzik1995,CieplakJCP2001,chanMIT}
For our present purpose, we consider two interaction matrices proposed 
recently for the study of IDP LLPS, namely,
$\varepsilon^{\rm KH}_{r,r'}$
obtained through a shifted MJ matrix,\cite{dignon18,MJ96} and
$\varepsilon^{\rm Mpipi}_{r,r'}$, which is partly based on explicit-water
simulations of pairwise amino acid residue interactions and has recently 
been used to successfully predict LLPS propensities of several biologically 
important IDPs\cite{Mpipi} (Fig.~1). The accuracies of truncated 
eigenvalue decompositions of
$\varepsilon^{\rm KH}_{r,r'}$ and $\varepsilon^{\rm Mpipi}_{r,r'}$ 
[Eq.~\eqref{eq:truncated}, with $\tilde a=1,2,3,4,$ and $5$] are
illustrated in Fig.~1a,b and the associated nonelectric charges 
associated with the $a=1,2,3,4$, and $5$ eigenvalues are provided in Fig.~1c.

\begin{figure}[!ht]
\centering
   \includegraphics[width=0.93\columnwidth]{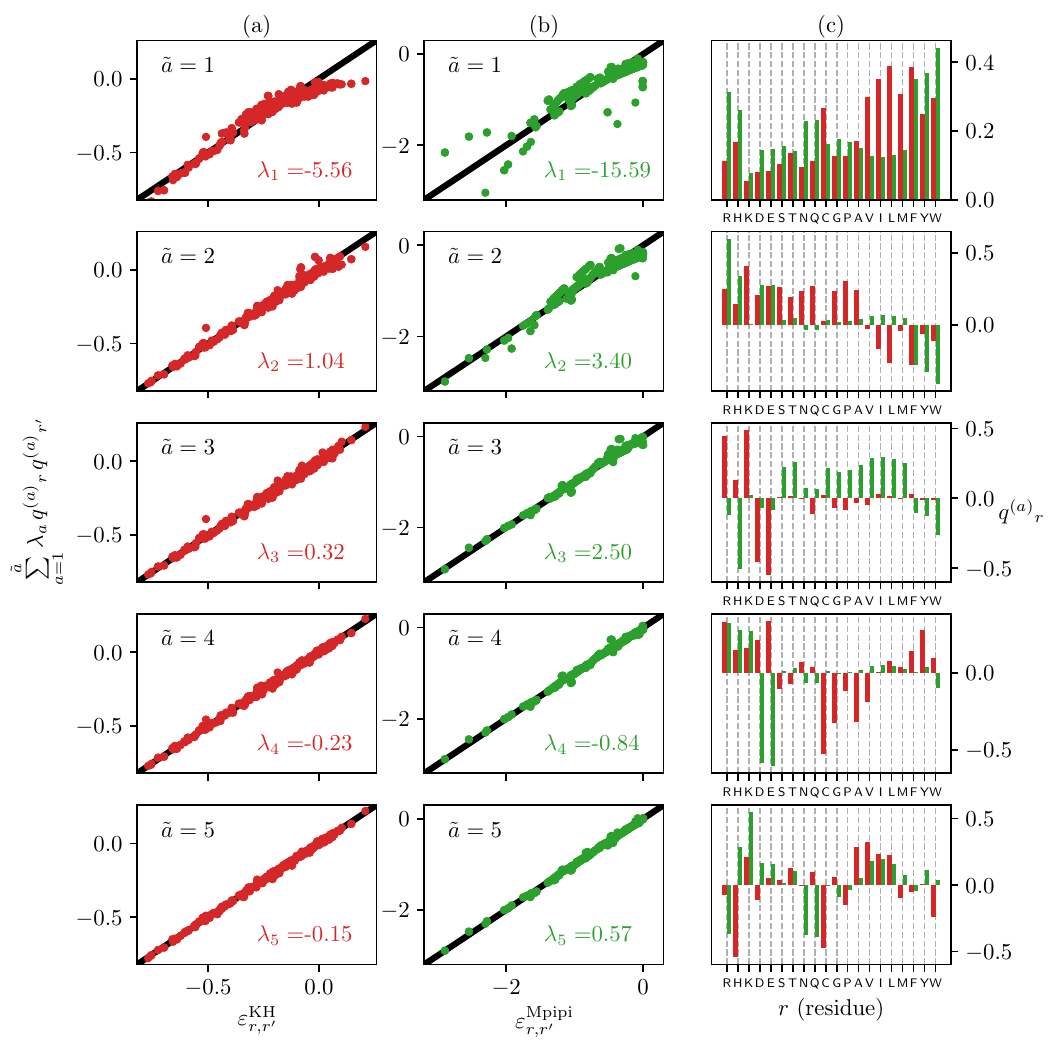}
\vskip -2 mm
\caption*{\footnotesize {\bf Fig.~1:} 
Eigenvalue decomposition of (a) $\varepsilon^{\rm KH}_{r,r'}$
and (b) $\varepsilon^{\rm Mpipi}_{r,r'}$ using
Eqs.~\eqref{eq:eps_spectral_decomposition} and \eqref{eq:truncated} where the
summations over eigenvalues are truncated at $\tilde{a}$.  Accuracy of a 
set of truncated summations (vertical variable) is indicated by
the proximity of the scatter plot [red (a) and green (b) data points]
to the slope $=1$ black line through the origin.  (c) Nonelectric
charges $q^{(a)}{}_{r}$, $a=1,2,3,4,$ and $5$ (top to bottom) of the amino acid
residue types (shown by their standard one-letter codes) are displayed for KH
(red) and Mpipi (green).  For KH, $\varepsilon^{\rm KH}_{r,r'}$ values
[horizontal variable in (a)] are taken from Table S3 of ref.~\citen{dignon18}
(corresponding to the ``KH-D'' model in this reference). For Mpipi,
$\varepsilon^{\rm Mpipi}_{r,r'}$ values [horizontal variable in (b)] are the
contact energies given in Supplementary Table~11 of ref.~\citen{Mpipi}.}
\label{ref:fig1}
\end{figure}


Fig.~1 shows significant deviations from $\varepsilon^{\rm KH}_{r,r'}$
or $\varepsilon^{\rm Mpipi}_{r,r'}$
when only one eigenvalue is used ($\tilde{a}=1$). At the same time, it
indicates that eigenvalue 
decomposition is reasonably accurate for $\tilde{a}\geq 2$. Interestingly, for 
$\tilde{a}=1$, i.e., when only a single nonelectric charge type is used to
describe the interactions,
large deviations are observed only for repulsive ($>0$) interactions in KH
but observed for strongly attractive (more negative) interactions as well
as a subset of the less attractive interactions in Mpipi.
When these two pairwise interaction schemes are compared, 
the nonelectric charges ($q^{(1)}{}_r$s) of the large 
nonpolar residues---valine, isoleucine, leucine, and methionine---associated 
with the energy matrix's most dominant ($a=1$) eigenvalue 
is much higher in KH than in Mpipi, reflecting the weaker interactions
among these nonpolar residues ascribed by Mpipi than by KH. But
the corresponding nonelectric charges for the 
aromatic residues phenylalanine, tyrosine, and tryptophan
are higher in Mpipi than in KH, reflecting the relatively more favorable
interactions enjoyed by these residues in Mpipi than in KH (see below).
As previously recognized,\cite{HaoLi1997,chan99,chanMIT} 
at least for KH, the nonelectric charge associated with the
most dominant eigenvalue corresponds to a measure of hydrophobicity,
with polar residues having smaller $q^{(1)}{}_r$s and large nonpolar 
residues having larger $q^{(1)}{}_r$s (red bars in the top panel of Fig.1c).
While all $a=1$ nonelectric charges are positive, 
$q^{(a)}{}_r$s can take positive and negative values for $a\geq 2$.
Intuitively, for KH, the nonelectric charge for $a=2$ corresponds to a measure
of polarity, with most polar residues having $q^{(2)}{}_r > 0$
and most nonpolar residues having $q^{(2)}{}_r < 0$; and the nonelectric
charge for $a=3$ appears to correlate strongly with electric charge, with 
arginine and lysine having the largest (most positive) $q^{(3)}{}_r$s whereas
aspartic and glutamic acids having the most negative $q^{(3)}{}_r$s 
(second and third panels from the top of Fig.1c). These observations
are tentalizing, but it should be emphasized that the nonelectric charges 
are basically mathematical constructs.
As such, they do not always lend themselves to simple interpretations
in terms of physical interaction types. A case in point is that the 
corresponding trend of $q^{(2)}{}_r$ and $q^{(3)}{}_r$ values in Mpipi is less 
straightforward to interpret in terms of a particular physical 
characteristics of the amino acid residues.
 
Besides KH and Mpipi, we consider three representative
interaction matrices constructed
from hydrophobicity scales that have recently been used for simulations 
of single-IDP conformations and biomolecular LLPS. We refer to these
interaction matrices as (i) $\varepsilon^{\rm HPS}_{r,r'}$, which is 
derived in ref.~\citen{dignon18} from an 
OPLS (optimized potentials for liquid simulations) forcefield-based\cite{} 
hydrophobicity scale,\cite{Rossky} 
(ii) $\varepsilon^{\rm Urry}_{r,r'}$, which is derived in
ref.~\citen{Urry-Mittal} from a prior analysis by Urry et al. of experimental 
data on heat-induced conformational compaction of host-guest 
polypentapeptides,\cite{Urry}
and (iii) $\varepsilon^{\rm FB}_{r,r'}$, which is derived in ref.~\citen{FB}
by optimizing agreement between coarse-grained molecular dynamics simulations
based on putative hydrophobicity scales
with experimental data on the radii of gyration of single-chain IDP 
conformational ensembles (a similar approach was also utilized in
ref.~\citen{krestenPNAS2021}).
These interaction matrices take the general form
\begin{equation}
\label{eq:eps_from_hydropathy}
\varepsilon^X_{r,r'} = \frac{\tilde{\lambda}_r + \tilde{\lambda}_{r'}}{2} 
+ \Delta , \quad X= {\rm HPS, Urry, FB},
\end{equation}
where $\tilde{\lambda}_r$ are twenty residue-specific 
hydrophobicity/hydropathy values [not to be confused with the 
eigenvalues $\lambda_a$ in Eq.~\eqref{eq:eps_spectral_decomposition}] 
and $\Delta$ is an overall
shift. Matrices in the form of Eq.~\eqref{eq:eps_from_hydropathy} have 
at most two nonzero eigenvalues, given by $\lambda_{1,2} = \frac {1}{2} 
\Bigl [ \sum_{r} \tilde{\lambda}_r + \Delta \pm 
\sqrt{20 \sum_{r} (\tilde{\lambda}_r + \Delta)^2} \; \Bigr ]$. 
A visual impression of the similarities and differences among
the five interaction matrices considered in this work are provided
by their depiction in Fig.~2 as heat maps.

\begin{figure}[!ht]
\centering
   \includegraphics[width=0.98\columnwidth]{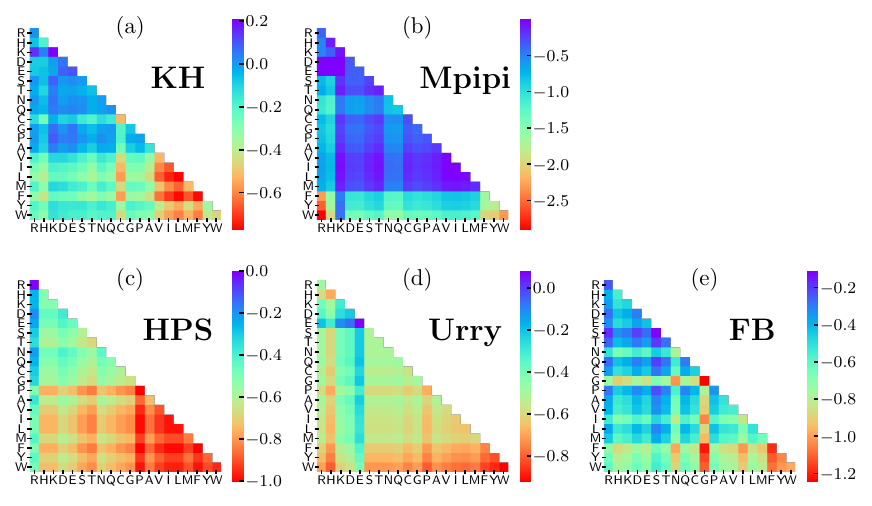}
\vskip -2 mm
\caption*{\footnotesize {\bf Fig.~2:} 
Heat-map representations of the interaction matrices considered in the
present work: (a) $\varepsilon^{\rm KH}_{r,r'}$,
(b) $\varepsilon^{\rm Mpipi}_{r,r'}$,  
(c) $\varepsilon^{\rm HPS}_{r,r'}$,
(d) $\varepsilon^{\rm Urry}_{r,r'}$, and 
(e) $\varepsilon^{\rm FB}_{r,r'}$. The
color scale ranges from red (most attractive) to blue (least
attractive or most repulsive). The HPS, Urry and FB matrices follow from
Eq.~\eqref{eq:eps_from_hydropathy} where the values of $\tilde{\lambda}_{r}$
are taken from Table~S1 in ref.~\citen{dignon18} for HPS, Table~S2 in
ref.~\citen{Urry-Mittal} for Urry, and Table~S7 in ref.~\citen{FB} for FB. 
Overall shift $\Delta = 0$ for both HPS and FB, 
whereas $\Delta = 0.08$ for Urry. 
Corresponding information for $\varepsilon^{\rm KH}_{r,r'}$ and 
$\varepsilon^{\rm Mpipi}_{r,r'}$ is provided in the caption of Fig.~1.
}
\label{ref:fig2}
\end{figure}

It follows from the above discussion that
formulation of the interactions entailed by HPS, Urry, and FB by eigenvalue
decomposition in terms of nonelectric charges are straightforward because 
it requires at most two types of nonelectric charges. By comparison,
the accuracy of corresponding formulations using a small number of 
types of nonelectric charges for interaction schemes encoded by 
$20\times 20$ energy matrices that are not derived from a $20$-value 
hydrophobicity scale (such as KH and Mpipi) has to be ascertained.
In Fig.~3, the accuracy of the truncated eigenvalue decompositions 
of $\varepsilon^{\rm KH}_{r,r'}$ and $\varepsilon^{\rm Mpipi}_{r,r'}$ is 
assessed by considering the quantity
\begin{equation}
\label{eq:Delta_eps}
\Delta \varepsilon_{r,r'} = \sum_{a=1}^{\tilde{a}} \lambda_a q^{(a)}{}_r q^{(a)}{}_{r'} - \varepsilon_{r,r'} , 
\end{equation}
which is the difference between the truncated 
summation and the original $\varepsilon_{r,r'}$. Results 
for KH and Mpipi at $\tilde{a}=1,2,$ and $3$ are shown, respectively, in 
Fig.~3a, b, and c.
The heat maps in Fig.~3 indicate that the KH interaction matrix is very 
well approximated at $\tilde{a}=3$ with the only exception of the
cysteine--cysteine (C--C) entry, which is noticeably overestimated at 
all $\tilde{a}\leq 3$. This exception is in line with the fact that 
the covalent C--C disulfide bond is physically different from all 
the other pairwise residue contacts that are noncovalent in nature. 
Disulfide bonds are significantly more important in folded proteins than
in IDPs. This may be a reason why the C--C exception in the accuracy of 
eigenvalue decomposition is observed in PDB-derived KH but not in the 
IDP physics/bioinformatics-derived Mpipi.
For Mpipi, the truncated eigenvalue 
decomposition at $\tilde{a}=3$ is reasonably accurate though it 
overestimates interactions involving the electrically negative aspartic 
and glutamic acid residues.

Under the next subheading, we will use the eigenvalue
decomposition in Eq.~\eqref{eq:eps_spectral_decomposition} to derive a
statistical field theory for protein solutions with spatially short-range
residue-dependent interactions governed by a matrix of contact energies
$\varepsilon_{r,r'}$. Relative to the field theory we consider
previously for polyampholytes,\cite{Pal2021,mimb2022,wessen2021,wessen2022}
the present extended field theory contains 
$\tilde{a}$ additional fields, i.e., one additional field $\varphi_{a}(\rr)$ 
for every eigenvalue $\lambda_a$ included, wherein
the field $\varphi_{a}(\rr)$ is conjugate to the density of nonelectric
charges $q^{(a)}{}_{r}$. 
Under subsequent subheadings, the new theory is first studied 
using an approximate analytical approach we find useful for computationally
efficient comparisons of interaction schemes encoded by different 
$\varepsilon_{r,r'}$s. The theory is then explored using FTS to take
into complete account of field fluctuations so as to provide, 
in the context of the present model, 
full sequence dependence of biomolecular LLPS and detailed 
structural information about residue partitioning inside protein-dense 
condensates. 

\begin{figure}[!t]
\centering
   \includegraphics[width=0.98\columnwidth]{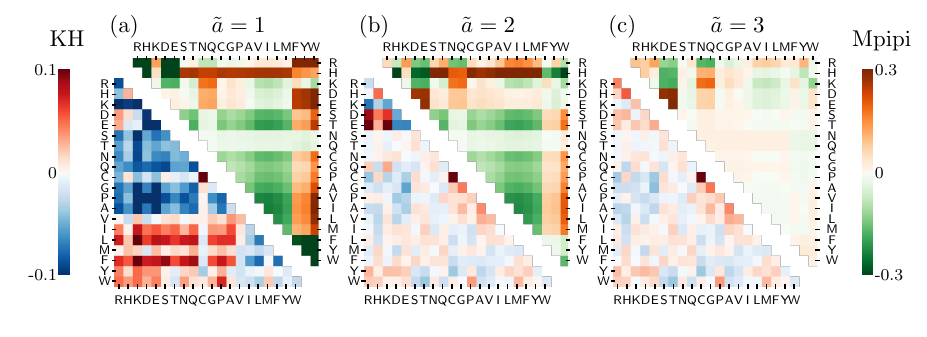}
\vskip -2 mm
\caption*{\footnotesize {\bf Fig.~3:} 
Accuracy of the truncated eigenvalue decomposition 
of $20\times 20$ interaction matrices that are not derived from
20-value hydrophobicity scales.
The level of accuracy for $\tilde{a}=1,2,$ and $3$ (a--c)
is quantified by $\Delta \varepsilon_{r,r'}$ in
Eq.~\eqref{eq:Delta_eps}. Bottom-left triangles using the blue--red
color scale on the left display $\Delta \varepsilon_{r,r'}$ for KH, whereas 
top-right triangles using the green-orange color scale on the right
display the same quantity for Mpipi.
}
\label{ref:fig3}
\end{figure}


{\bf Statistical Field Theory.}
Here we consider a system of volume $V$ with $\np$ protein chains, each
consisting of $N$ consecutive amino acid residues of types $r_{\alpha}$ (with
$\alpha=1,\dots,N$), and denote the position of residue $\alpha$ on chain $i$
($=1,\dots,\np$) as $\RR_{i,\alpha}$. The system contains also $n_{+}$
unit-positive and $n_{-}$ unit-negative ions with positions $\rr_{\pm, i}$, and
$\nw$ neutral solvent particles with positions $\rr_{{\rm w},i}$. We let
$\rho_{\rm p,\pm,w} \equiv n_{\rm p,\pm,w} / V$ denote the bulk number
densities of the polymers, ions and solvent in the system, 
and $\rho_{\rm b} = Nn_{\rm p}/V = N\rho_{\rm p}$ the
bulk number density of polymer beads. The canonical partition function
of this system is given by
\begin{equation} \label{eq:partition_func_particle}
Z = \frac{1}{\np! n_+! n_-! \nw!} \int \lbrace \d \RR \rbrace \lbrace \d \rr_{+} \rbrace \lbrace \d \rr_{-} \rbrace \lbrace \d \rr_{\rm w} \rbrace \, \e^{-\hat{H} } , 
\end{equation}
where
\begin{equation}
\lbrace \d \RR \rbrace \equiv \prod_{i=1}^{\np} \prod_{\alpha=1}^N \d
\RR_{i,\alpha} \, , \quad \lbrace \d \rr_{\pm} \rbrace \equiv
\prod_{i=1}^{n_{\pm}}  \d \rr_{\pm,i} \, , \quad \lbrace \d \rr_{\rm w} \rbrace
\equiv \prod_{i=1}^{\nw} \d \rr_{ {\rm w},i} \, 
\end{equation}
are position integration measures for polymer beads, ions 
and solvents, respectively. The microscopic Hamiltonian in units
of $k_{\rm B}T$, where
$k_{\rm B}$ is Boltzmann's constant and $T$ is absolute temperature,
written as
\begin{equation} \label{eq:particle_H}
\hat{H} = \hat{H}_0 + \hat{H}_{\rm c} + \hat{H}_{\rm e} + \hat{H}_{\rm h} \; ,
\end{equation}
contains terms accounting for chain connectivity ($\hat{H}_0$), 
soft compressibility ($\hat{H}_{\rm c}$), electrostatic interactions 
($\hat{H}_{\rm e} $) and spatially short-range 
hydrophobic-like interactions---which may include van der Waals, 
$\pi$-related, and other forms of short-spatial-range interactions
($\hat{H}_{\rm h}$). The chain connectivity term
\begin{equation}
\hat{H}_0 = \frac{3}{2 b^2} \sum_{i=1}^{\np} \sum_{\alpha=1}^{N-1}
(\RR_{i,\alpha+1} - \RR_{i,\alpha})^2 
\end{equation}
is controlled by the segment length $b$ which we set to the trans
C$_\alpha$--C$_\alpha$ virtual bond length, i.e., $b=3.8$ {\AA}, 
throughout this work. The soft-compressibility term
\begin{equation}
\label{eq:Hc}
\hat{H}_{\rm c} = \frac{1}{2\gamma} \int \d \rr \left[ \hat{\rho}(\rr) - \rho_0
\right]^2 
\end{equation}
penalizes deviations of the total density 
\begin{equation}
\hat{\rho}(\rr) = v_{\rm b} \hat{\rho}_{\rm b}(\rr) + v_{+}
\hat{\rho}_{+}(\rr) + v_{-} \hat{\rho}_{-}(\rr) + v_{\rm w} 
\hat{\rho}_{\rm w}(\rr) 
\end{equation}
from a reference density $\rho_0$, and thus serves to capture an essential
aspect of excluded volume effects.
In the above equation, $\hat{\rho}_i(\rr)$
with $i={\rm b}, +, -, {\rm w}$ (referring to polymer bead, cation, anion, and
solvent particle, respectively) is the number density of species $i$ at
position $\rr$ and the $v_i$s are factors that we use to model the relative
volume of the different particles. The number densities are
\begin{equation}
\label{eq:Gamma}
\hat{\rho}_{\rm b}(\rr) = \sum_{i=1}^{\np} \sum_{\alpha = 1}^N \Gamma(\rr -
\RR_{i,\alpha}) \quad \mbox{and} \quad \hat{\rho}_i(\rr) = \sum_{j=1}^{n_i}
\Gamma(\rr - \rr_{i,j}) ,\quad  i=+,-,{\rm w} ,
\end{equation}
where individual particles are modeled as Gaussian distributions $\Gamma(\rr)
= \e^{-\rr^2 / 2 a_{\rm s}^2} / (2 \pi a_{\rm s}^2)^{3/2}$, which serve to 
regularize infinities arising from particle contact interactions and self 
interactions.\cite{Wang2010,Riggleman2012}
Here the smearing parameter $a_{\rm s}$ is equal to $1/\sqrt{3}$ of
the standard deviation of ${\bf r}$ in the $\Gamma(\rr)$ distribution
(i.e., $a_{\rm s}=\sqrt{\langle |\bm{r}|^2\rangle/3}$). 
We refer to $\gamma$ in Eq.~\eqref{eq:Hc}---which has units of 
density---as the compressibility, and note that in the limit of
$\gamma\rightarrow 0$, we obtain the standard incompressibility 
condition $\e^{-\hat{H}_{\rm c}} \propto \delta[\hat{\rho}(\rr) - \rho_0]$ 
where the functional $\delta$-function enforces $\hat{\rho}(\rr) = \rho_0$ 
at all spatial positions $\rr$. 

Electrostatic interactions are included in $\hat{H}_{\rm e}$ through 
the standard pairwise Coulomb potential,
\begin{equation}
\hat{H}_{\rm e} = \frac{ \lB }{2} \int \d \rr \int \d \rr' \, \frac{\hat{c}(\rr) \hat{c}(\rr') }{|\rr - \rr'|} \, , 
\end{equation}
where
\begin{equation}
\label{eq:hatc}
\hat{c}(\rr) = \sum_{i=1}^{\np} \sum_{\alpha=1}^N \sigma_{i,\alpha} \Gamma(\rr - \RR_{i,\alpha}) + \hat{\rho}_+(\rr) - \hat{\rho}_-(\rr) 
\end{equation}
is the density of electric charge at $\rr$ and $\sigma_{i,\alpha}$ is the
electric charge of the $\alpha$th residue on the $i$th chain 
in units of the proton charge $e$. Because the present study is limited
to systems of chains with identical sequence, i.e., 
$\sigma_{i,\alpha}=\sigma_{\alpha}$ is independent of $i$, 
the subscript $i$ will be dropped from the symbol $\sigma$ for
electric charge hereafter. The
strength of electrostatic interactions is controlled by the Bjerrum length $\lB
= e^2/4\pi\epsilon_0\epsilon_{\rm r}k_{\rm B}T$ that involves vacuum
permittivity $\epsilon_0$ and a constant background 
relative permittivity 
$\epsilon_{\rm r}$. In this work, we neglect any concentration dependence
of $\epsilon_{\rm r}$ as this has been shown to amount to only small to 
moderate effects on LLPS propensity.\cite{wessen2021} Nonetheless, if 
desired, it would be straightforward to implement a concentration dependent 
permittivity in our model through the approach described in 
ref.~\citen{wessen2021}. 

The final term in the Hamiltonian in Eq.~\eqref{eq:particle_H}, 
$\hat{H}_{\rm h}$, contains the pairwise residue-specific spatially 
short-range interactions, which we express as 
\begin{equation} \label{eq:H_hat_pos}
\hat{H}_{\rm h} = \frac{1}{2} \sum_{i,j=1}^{\np} \sum_{\alpha,\beta=1}^N
\varepsilon_{r_{\alpha}, r_{\beta}} \, \Gamma^2 \star V_{\rm h}(|\RR_{i,\alpha}
- \RR_{j,\beta}|) 
\end{equation}
where
\begin{equation}
\Gamma^2 \star V_{\rm h}(|\RR_{i,\alpha} - \RR_{j,\beta}|) \equiv \int \d \rr
\int \d \rr' \, \Gamma(\rr-\RR_{i,\alpha}) \Gamma(\rr' - \RR_{j,\beta}) 
V_{\rm h}(|\rr - \rr'|) 
\end{equation}
accounts for the residue-residue interaction in the presence of Gaussian 
smearing (note that $\Gamma^2 \star V_{\rm h} \rightarrow V_{\rm h}$ as 
$a_{\rm s} \rightarrow 0$). Leaving the functional form of the interaction 
potential $V_{\rm h}(|\rr|)$ (in units of $k_{\rm B}T$) 
unspecified for the moment, we have in the above 
expression assumed that all residue dependence is accounted for by the 
overall multiplicative factors $\varepsilon_{ r_{\alpha},r_{\beta} }$
from a given interaction matrix.
The spectral form of $\varepsilon_{r,r'}$ in
Eq.~\eqref{eq:eps_spectral_decomposition} can be used to write 
$\hat{H}_{\rm h}$ in Eq.~\eqref{eq:H_hat_pos} as
\begin{equation}
\hat{H}_{\rm h} = \sum_{a=1}^{20} \frac{1}{2} \int \d \rr \int \d \rr' \, 
V_{\rm h}(|\rr - \rr'|)  \lambda_a \hat{h}_a(\rr) \hat{h}_a(\rr'), 
\end{equation}
where 
\begin{equation}
\hat{h}_a(\rr) = \sum_{i=1}^{\np} \sum_{\alpha=1}^{N} q^{(a)}{}_{r_{\alpha}}
\Gamma(\rr - \RR_{i,\alpha}) 
\end{equation}
is the density of nonelectric charge for the $a$th eigenvalue $\lambda_a$ 
at position $\rr$.

The Hamiltonian terms $\hat{H}_{\rm c, e, h}$ for non-bonded interactions are
quadratic in densities and can thus be decoupled using standard
Hubbard-Stratonovich transformations,\cite{Fredrickson2006} turning the
partition function in Eq.~\eqref{eq:partition_func_particle} into that of a
statistical field theory. Special care needs to be given to $\hat{H}_{\rm h}$
wherein terms with $\lambda_a > 0$ and $\lambda_a<0$ need to be linearized with
fields integrated on the imaginary and real axis, respectively. To keep track
of the sign of eigenvalue $\lambda_a$, we introduce the variable $\xi_a$,
defined as
\begin{equation}
\xi_a = \left\lbrace \begin{matrix}
-\i & , & \lambda_a < 0 \\
1 &, & \lambda_a > 0
\end{matrix} \right. , 
\end{equation}
where $\i^2=-1$.
The partition function $Z$ in Eq.~\eqref{eq:partition_func_particle} can then
be shown to be equivalent to a statistical field theory with partition function
\begin{equation} \label{eq:partition_func_field}
Z = \frac{V^{\np + n_+ + n_- + \nw} }{\np! n_+! n_-! \nw!} \int \D \eta \int \D
\psi \left( \prod_{a=1}^{20} \int \D \varphi_a \right) \e^{-H[\eta,\psi,\lbrace
\varphi_a \rbrace]} ,
\end{equation}
where the field Hamiltonian (in units of $k_{\rm B}T$)
\begin{equation} \label{eq:field_hamiltonian}
H = - \sum_{i={\rm p, +, -, w}} n_i \ln Q_i[\breve{\eta},\breve{\psi},\lbrace
\breve{\varphi}_a \rbrace] + \int \d \rr \left[  - \i \rho_0 \eta + \frac{
\gamma \eta^2}{2}+ \frac{(\bm{\nabla} \psi)^2}{8 \pi \lB} + \sum_{a=1}^{20}
\frac{ \varphi_a V_{\rm h}{}^{-1} \varphi_a }{2 |\lambda_a | } \right] . 
\end{equation}
In the field theory representation of the system, $\eta(\rr)$ and $\psi(\rr)$
are conjugate fields to the density $\hat{\rho}(\rr)$ and the electric charge
density $\hat{c}(\rr)$, respectively, while $\varphi_a(\rr)$ is conjugate to
the nonelectric charge density $\hat{h}_a(\rr)$ associated with eigenvalue
$\lambda_a$ in the spectral decomposition of the interaction matrix. 
The single-molecule
partition functions $Q_{\rm p, \pm, w}[\breve{\eta},\breve{\psi},\lbrace
\breve{\varphi}_a \rbrace]$ depend on smeared fields $\breve{\phi}(\rr) =
\Gamma \star \phi(\rr) \equiv \int \d \rr' \Gamma(\rr - \rr') \phi(\rr')$ (for
$\phi=\eta,\psi,\varphi_a$) via the relations
\begin{eqnarray}
\label{eq:Qpm}
Q_{\pm} &=& \frac{1}{V} \int \d \rr \exp\left[ - \i ( v_{\pm} \breve{\eta}(\rr)
\pm \breve{\psi}(\rr)) \right] \, , \\
\label{eq:Qw}
Q_{\rm w} &=& \frac{1}{V} \int \d \rr \exp\left[ - \i v_{\rm w}
\breve{\eta}(\rr) \right] \, 
\end{eqnarray}
for ions and solvents, respectively, and
\begin{equation}
\label{eq:Qp}
Q_{\rm p} = \frac{1}{\mathcal{N}} \left( \prod_{\alpha=1}^N \int \d
\RR_{\alpha} \right) \exp\left[ - \frac{3}{2b^2}\sum_{\alpha=1}^{N-1}
(\RR_{\alpha+1} - \RR_{\alpha})^2 - \sum_{\alpha=1}^N W_{\alpha}(\RR_{\alpha})
\right] \, 
\end{equation}
for polymer chains, where $\mathcal{N}\equiv V (2 \pi b^2 / 3)^{3 (N-1)/2}$ normalizes $Q_{\rm p}$ to unity at zero field values, and 
\begin{equation}
\label{eq:Wa}
W_{\alpha}(\rr) = \i v_{\rm b} \breve{\eta}(\rr) + \i \sigma_{\alpha} \breve{\psi}(\rr) + \sum_{a=1}^{20} \i \, \xi_a q^{(a)}{}_{r_{\alpha}} \breve{\varphi}_a(\rr) \, .
\end{equation}
In Eq.~\eqref{eq:field_hamiltonian}, $V_{\rm h}{}^{-1}$ represents 
a differential operator satisfying 
\begin{equation}
\label{eq:Vh}
V_{\rm h}{}^{-1} V_{\rm h}(|\rr|) = \delta(\rr) 
\end{equation} 
and thus $V_{\rm h}{}^{-1}$ depends on the functional form of 
$V_{\rm h}(|\rr|)$. 
Since our primary goal here is formulation development,
for simplicity we consider only temperature-independent forms
of the spatially short-range interaction potential $V_{\rm h}(|\rr|)$.
Physically, however, solvent-mediated interactions of amino acid residues 
can be temperature dependent when viewed in an implicit-solvent
perspective,\cite{biochemrev} 
especially for the hydrophobic effects among
nonpolar residues,\cite{dillchan97}
leading, e.g., to cold denaturation\cite{Dill1989} and 
LLPS driven by increasing
temperature rather than decreasing temperature.\cite{Roland2019}
When necessary,
these effects can be readily incorporated into our formulation by
introducing temperature-dependent 
potentials,\cite{dillchan97,Dill1989,Kaya2003c} i.e.,
$V_{\rm h}(|\rr|)\rightarrow V_{\rm h}(|\rr|,T)$ 
as in ref.~\citen{Mittal-ACSCent2019}.
\\



{\bf Analytical Calculation of the Leading Order Free Energy.}
We next proceed to an approximate evaluation of the statistical field theory 
partition function $Z$ in Eq.~\eqref{eq:partition_func_field}. In the
analytical approach taken here, we only account for the leading-order
effects from each field around the homogeneous saddle solution. For our
model, this amounts to the mean field theory (MFT) solution for the fields
$\eta$ and $\varphi_a$, while electrostatic effects first show up at the
one-loop level in $\psi$ which are accounted for by RPA.


The spatially homogeneous MFT solution is obtained by first letting the fields
$\eta$, $\psi$ and $\varphi_a$ in the expression for the field Hamiltonian $H$
in Eq.~\eqref{eq:field_hamiltonian} be position-independent 
(hence smeared fields $\breve{\phi}=\phi$
for $\phi=\eta,\psi,\varphi_a$ are also position-independent) and then
requiring that the first derivatives of the resulting $H$ with respect
to the fields are zero. With position-independent fields, the 
single-molecule partition functions $Q_i$, defined in 
Eqs.~\eqref{eq:Qpm}---\eqref{eq:Qp}, become
\begin{equation}
\label{eq:Q-pi}
\begin{aligned}
Q_{\rm w} &= \e^{-\i v_{\rm w} \eta} , \\
Q_{\pm} &= \e^{-\i v_{\pm} \eta \mp \i \psi} , \\
Q_{\rm p} &= \exp\left[ -\i N v_{\rm b} \eta - \i \sigma_{\rm p} \psi - \i \sum_{a=1}^{20} \xi_a q^{(a)} \varphi_a \right] ,
\end{aligned}
\end{equation}
where $\sigma_{\rm p} = \sum_{\alpha=1}^N \sigma_{\alpha}$ is the net charge
of a polymer and $q^{(a)} = \sum_{\alpha=1}^N q^{(a)}{}_{r_\alpha}$ is 
the total nonelectric charge associated with the $a$th eigenvalue of 
one polymer chain. The above relations for $Q_i$ lead to the following 
expression for $H$ with homogeneous fields:
\begin{equation} \label{eq:H_const_fields}
\begin{aligned}
H =& \i ( n_{\rm p} N n_{\rm b} + n_{+} v_{+} + n_{-} v_{-} + n_{\rm w} 
v_{\rm w} ) \eta + \i (n_{\rm p} \sigma_{\rm p} + n_{+} - n_{-} ) \psi + 
\i n_{\rm p}
\sum_{a=1}^{20} \xi_a q^{(a)} \varphi_a \\ &+  V \left[ - \i \rho_0 \eta +
\frac{\gamma \eta^2}{2} + \sum_{a=1}^{20} \frac{\varphi_a^2}{2 |\lambda_a|
\hat{V}_{\rm h}(\bm{0})} \right].
\end{aligned}
\end{equation}
In Eq.~\eqref{eq:H_const_fields}, $\hat{V}_{\rm h}(\bm{0})$ is the 
$\kk=\bm{0}$ component of the Fourier transform 
$\hat{V}_{\rm h}(\kk) = \int \d \rr \, \e^{-\i \rr \cdot \kk} V_{\rm h}(|\rr|)$ 
of the spatially short-range potential $V_{\rm h}(|\rr|)$.
This factor can be derived by considering the Fourier representation 
of $\delta(\rr) = (2 \pi)^{-3} \int \d \kk \, \exp(\i \kk \cdot \rr)$ 
in the defining Eq.~\eqref{eq:Vh} 
for $V_{\rm h}^{-1}$, 
which leads to $V_{\rm h}^{-1} \exp(\i \kk \cdot \rr) = 
\exp(\i \kk \cdot \rr) / \hat{V}_{\rm h}(\kk)$. Then, by noting that 
the Fourier transform of the position-independent $\varphi_a$ is proportional 
to $\delta(\kk)$, it is straightforward to show that the $\int \d \rr$ 
integration with position-independent $\varphi_a$ produces the 
$ \hat{V}_{\rm h}(\bm{0}) $ term.

The coefficient of $\i \psi$ in Eq.~\eqref{eq:H_const_fields} is the 
total electric charge in the system which must be set to zero, 
$n_{\rm p} \sigma_{\rm p} + n_{+} - n_{-} = 0$. The charge-conjugate 
field $\psi$ thus vanishes in Eq.~\eqref{eq:H_const_fields} with no
contribution to the MFT solution. For the other fields, solving 
$\partial H / \partial \eta = \partial H / \partial \varphi_a = 0$ 
for $\bar{\eta}$ and $\bar{\varphi}_a$ leads to
\begin{equation} \label{eq:MFT_solution}
\bar{\eta} = \frac{-\i \Delta\rho}{\gamma} \quad \mbox{and} \quad \bar{\varphi}_a = -\i \hat{V}_{\rm h}(\bm{0}) \xi_a |\lambda_a| q^{(a)} \rhop ,
\end{equation}
where $\Delta \rho \equiv \nu_{\rm b} N \rhop + \nu_+ \rho_+ + \nu_- \rho_- +
\nu_{\rm w} \rho_{\rm w} - \rho_0$ is the deviation of the total bulk density
from $\rho_0$, and $q^{(a)} \equiv \sum_{\alpha=1}^N q^{(a)}{}_{r_\alpha}$ is
the total nonelectric charge associated with the $a$th eigenvalue $\lambda_a$ 
of one
polymer chain. The same expressions for $\bar{\eta}$ and $\bar{\varphi}_a$
[Eq.~\eqref{eq:MFT_solution}] may alternatively be obtained by requiring
vanishing first functional derivatives of $H$ with respect to
$\eta(\rr)$ and $\varphi_a(\rr)$ and
then seeking position-independent solutions $\eta(\rr) \rightarrow \bar{\eta}$,
$\psi(\rr) \rightarrow \bar{\psi}$ and $\varphi_a(\rr) \rightarrow
\bar{\varphi}_a$ of the resulting $\delta H / \delta \eta(\rr) 
= \delta H / \delta \psi(\rr) = \delta H / \delta \varphi_a(\rr) = 0$ 
conditions, as will be discussed below under the next subheading.

Plugging the MFT solution Eq.~\eqref{eq:MFT_solution} back into 
Eq.~\eqref{eq:H_const_fields} leads to the field Hamiltonian per 
unit volume $\bar{h}\equiv H/V$, given by 
\begin{equation}
\label{eq:barh}
\bar{h} = \frac{{\Delta \rho}^2}{2 \gamma} + \frac{\hat{V}_{\rm h}(\bm{0})}{2}
\left[ \sum_{a=1}^{20} \lambda_a ({q^{(a)}})^2 \right] \rhop^2 \; .  
\end{equation}
In the approximate analytical approach taken in this work, we set $\eta(\rr)$
and $\varphi_a(\rr)$ to their MFT values $\bar{\eta}$ and $\bar{\varphi}_a$,
while a spatially varying $\psi(\rr)$ is kept to $\mathcal{O}(\psi^2)$ in $H$.
The field Hamiltonian in this approximation is therefore $H \approx V \bar{h} +
H_{\rm RPA}[\psi]$ where the $\mathcal{O}(\psi^2)$ terms are contained in
\begin{equation}
H_{\rm RPA} = \int\frac{\d \kk}{(2 \pi)^3} \frac{1}{2} \hat{\psi}(-\kk) \left[ \hat{\Gamma}^2 \left( N \rhop  g_{\rm cc}(|\kk|) + \rho_+ + \rho_- \right) + \frac{\kk^2}{4 \pi \lB} \right] \hat{\psi}(\kk) .
\end{equation}
The factor $g_{\rm cc}(k) $, given by\cite{wessen2021}
\begin{equation}
g_{\rm cc}(k) = \frac{1}{N}\sum_{\alpha,\beta=1}^N \sigma_{\alpha}
\sigma_{\beta} \e^{-|\alpha-\beta| b^2 k^2 / 6} ,
\end{equation}
comes from the quadratic expansion of the single polymer partition 
function $Q_{\rm p}$. In our approximation, the functional integrals 
in the partition function $Z$ amounts to
\begin{equation} \label{eq:full_to_MFT}
\int \D \eta \int \D \psi \left( \prod_{a=1}^{20} \int \D \varphi_a\right)
\e^{-H[\eta, \psi,\lbrace \varphi_a \rbrace]} \approx \e^{- V \bar{h}} \int \D
\psi \, \e^{-H_{\rm RPA}} 
\end{equation}
where $V \bar{h} $ is the field Hamiltonian evaluated at homogeneous 
field values in accordance with Eq.~\eqref{eq:barh}.


To further simplify the formulation, 
we now take the $\gamma \rightarrow 0$ limit
such that solvent density is no longer an independent component of the system 
but is instead determined by 
\begin{equation}
\rho_{\rm w} = \frac{1}{v_{\rm w}} \left( \rho_0 - v_{\rm b} N \rho_{\rm p} - 
v_+ \rho_+ - v_- \rho_- \right) .
\end{equation}
In other words, $\Delta \rho = 0$ and, equivalently, 
$\rho_0=v_{\rm b} N \rho_{\rm p} +
v_+ \rho_+ + v_- \rho_- + v_{\rm w}\rho_{\rm w}$. 
The remaining system components 
$\lbrace \rhop, \rho_+, \rho_- \rbrace$ are further constrained by
electric charge neutrality of the system as a whole, 
i.e.~$\sigma_{\rm p} \rhop + \rho_+ - \rho_- = 0$, where
$\sigma_{\rm p}$ is the net electric charge of a polymer chain.
To make this condition
manifest, we introduce the overall salt density $\rho_{\rm s}$, which is the
overall number density of cation-anion pairs. The ion type in excess is
referred to as counterions. For net-neutral or
net-positive chains, $\sigma_{\rm p}\geq 0$, we have $\rho_+ = \rho_{\rm s}$
and $\rho_- = \sigma_{\rm p} \rho_{\rm p} + \rho_{\rm s}$;
for net-negative chains, $\sigma_{\rm p} < 0$, we have
$\rho_- = \rho_{\rm s}$
and $\rho_+ = -\sigma_{\rm p} \rho_{\rm p} + \rho_{\rm s}$.
Taking all the above considerations together and with $k\equiv |\kk|$,
the MFT/RPA free energy per volume in units of $k_{\rm B}T$, 
$f(\rhop,\rho_{\rm s})\equiv - (\ln Z)/V$, becomes
\begin{equation} \label{eq:MFT_free_energy}
f(\rhop, \rho_{\rm s}) = - s(\rhop,\rho_{\rm s}) - \rho_{0} \chi_{\rm eff}
\phi_{\rm p}^2 
+ \frac{1}{4 \pi^2} \int_0^{\infty} \d k \, k^2 \ln\left[ 1 + \frac{4 \pi
\lB}{k^2} \hat{\Gamma}^2 \left( 2 \rho_{\rm s} + |\sigma_{\rm p}| \rhop +
g_{\rm cc}(k) \rho_{\rm b} \right) \right] 
\end{equation}
after performing the $\int \D \psi$ integration\cite{wessen2021} 
in Eq.~\eqref{eq:full_to_MFT}, wherein the term
\begin{equation}
-s(\rho_{\rm p}, \rho_{\rm s}) 
= \sum_{i = {\rm p, +, -, w}} \rho_i \ln \rho_i \, 
\end{equation}
accounts for translational entropy and follows from applying Stirling's
approximation $\ln n! \approx n \ln n - n$ to the factorial prefactors in
Eq.~\eqref{eq:partition_func_field} and neglecting terms linear in $\rho_i$ 
arising from the nonlogarithmic $-n$ part of the Stirling approximation
because terms linear in $\rho_i$ in the free energy have no effect 
on phase separation properties.\cite{linJML}
In the second term on the right hand side of 
Eq.~\eqref{eq:MFT_free_energy}, $\phi_{\rm p} \equiv v_{\rm b} N \rho_{\rm p} /
\rho_0$ is the volume fraction occupied by polymer beads (and thus the
volume fraction of the polymers themselves) and $\chi_{\rm eff}$
is a dimensionless effective Flory-Huggins $\chi$-parameter that originates
from the spatially short-range interactions [cf. second 
term of $\bar{h}$ in Eq.~\eqref{eq:barh}]:
\begin{equation} \label{eq:chi_eff}
\chi_{\rm eff} = - \frac{ \rho_0 \hat{V}_{\rm h}(\bm{0}) }{2 v_{\rm b}^2 N^2}
\sum_{a=1}^{20} \lambda_a ({q^{(a)}})^2 \, . 
\end{equation}
Using Eq.~\eqref{eq:eps_spectral_decomposition}, 
the summation over $a$ in the above expression for $\chi_{\rm eff}$ 
can be reverted, i.e., 
\begin{equation}
\label{eq:lambda-sum}
\sum_{a=1}^{20} \lambda_a ({q^{(a)}})^2 = \sum_{\alpha=1}^N \sum_{\beta=1}^N
\varepsilon_{r_{\alpha}, r_{\beta}} \; ,
\end{equation}
to show that it is a summation over all possible
pairwise residue-residue interaction energies
for a pair of heteropolymer chains with the given sequence of residues.
As such, the present mean-field treatment of spatially short-range
interactions is akin to the 
random-mixing or Bragg-Williams\cite{bragg-williams} approximation
used in Flory-Huggins theories of proteinlike 
heteropolymers.\cite{Dill1989,dill1985}
As mentioned above, the third term
in Eq.~\eqref{eq:MFT_free_energy} corresponds to the standard RPA term
accounting for Gaussian fluctuations in electric charge density. This
term is obtained from functional integral over $\psi$ in 
Eq.~\eqref{eq:full_to_MFT} followed by subtraction of the free-energy 
contribution at $\rhop = \rho_{\rm s} = 0$, a subtraction that has no
effect on phase separation properties. Given the free energy 
$f(\rhop, \rho_{\rm s})$ in Eq.~\eqref{eq:MFT_free_energy}, phase 
diagrams may be constructed, for example, by matching the polymer and 
salt chemical potentials $\mu_{\rm p,s} = 
\partial f / \partial \rho_{\rm p,s}$ and osmotic pressure
$\Pi = \mu_{\rm p} \rho_{\rm p} + \mu_{\rm s} \rho_{\rm s} - f$ 
(ref.~\citen{mimb2022}).
\\


{\bf Field-Theoretic Simulations (FTS).}
One shortcoming of the above analytical approximation is that all
effects of the spatially short-range interactions are condensed 
into the value of $\chi_{\rm eff}$ which depends only on overall 
residue composition of the polymer but not the specific sequential
arrangement of the residues. To tackle such sequence-specific
effects, one needs to account for higher-order fluctuations in the partition
function in Eq.~\eqref{eq:partition_func_field}. 
To this end, we next consider ways to
study systems described by Eq.~\eqref{eq:partition_func_field} using 
FTS because of its ability to afford, in principle, a full account of
field fluctuations.\cite{Fredrickson2002}

In FTS, each field (denoted generically as $\phi$) 
is analytically continued into its complex plane, and
approximated by a set of discrete variables defined on the sites of a cubic
$M^3$ lattice with periodic boundary conditions and lattice spacing $\Delta x$.
The fields evolve in a fictitious complex-Langevin (CL) time $t$ according to
\begin{equation} \label{eq:CL_evolution}
\frac{\partial \phi(\rr,t)}{\partial t} = - \frac{\delta H}{\delta \phi(\rr,t)} + \tilde{\eta}_{\phi}(\rr,t) \quad , \quad \phi = \eta, \psi, \varphi_a ,
\end{equation}
where $\tilde{\eta}_{\phi}(\rr,t)$ represents real-valued Gaussian noise with
zero mean, i.e., 
$\langle \tilde{\eta}_{\phi}(\rr,t) \tilde{\eta}_{\phi'}(\rr',t')
\rangle = 2\delta_{\phi,\phi'} \delta(t-t') \delta(\rr - \rr')$,
where $\delta_{\phi,\phi'}=1$ if
$\phi=\phi'$ and $\delta_{\phi,\phi'}=0$ if $\phi\neq\phi'$. Thermal
averages of the real-space system can then be computed as 
asymptotic CL time averages of field operators in the fictitious-time 
system that are constructed to correspond to the thermodynamic 
observable of interest.\cite{Parisi1983,Klauder1983} 
This approach has its origin in the development, beginning in the
1980s, of stochastic quantization\cite{ParisiWu1981}
as a method for studying quantum field theories and their
regularization\cite{HSCMartyGhost,HSCMartyGravity,rumpf} 
and has since been applied extensively to study properties of
polymer solutions.\cite{Fredrickson2006}

In the present model, the functional derivatives of the field 
Hamiltonian $H$ in Eq.~\eqref{eq:field_hamiltonian} are
\begin{equation}\label{eq:H_derivatives}
\begin{aligned}
\frac{\delta H}{\delta \eta(\rr) } &= \i \sum_{i={\rm, b,\pm, w}} 
v_i \tilde{\rho}_i(\rr) -  \i \rho_0 + \gamma \eta(\rr) , \\
\frac{\delta H}{\delta \psi(\rr) } &= \i  \tilde{c}(\rr) - \frac{\bm{\nabla}^2 \psi(\rr)}{4 \pi \lB} , \\
\frac{\delta H}{\delta \varphi_a(\rr) } &= \i  \xi_a \tilde{h}_a(\rr) +
\frac{1}{|\lambda_a|} V_{\rm h}^{-1} \varphi_a(\rr) \; ,
\end{aligned}
\end{equation}
where $\tilde{\rho}_i(\rr)$ in the first relation here 
in Eq.~\eqref{eq:H_derivatives}
is a field operator corresponding to the 
density of component $i$, and is given, respectively, by
\begin{equation}
\label{eq:rho-ops}
\tilde{\rho}_{\pm}(\rr) =  \Gamma \star \frac{\rho_{\pm} }{Q_{\pm}} \e^{-\i
v_{\pm} \breve{\eta}(\rr) \mp \i \breve{\psi}(\rr)} \quad \mbox{and} \quad
\tilde{\rho}_{\rm w}(\rr) =  \Gamma \star \frac{\rho_{\rm w} }{Q_{\rm w}}
\e^{-\i v_{\rm w} \breve{\eta}(\rr) }  
\end{equation}
for ions and solvents, whereas the polymer bead density operator, 
$\tilde{\rho}_{\rm b}(\rr)$, is computed through forward and backward 
chain propagators $q_{\rm F,B}(\rr,\alpha)$ which are constructed
by applying the following relations 
iteratively:\cite{Pal2021,mimb2022,wessen2021}
\begin{equation}
\begin{aligned}
q_{\rm F}(\rr, \alpha+1) = \e^{- W_{\alpha+1}(\rr)} \, 
\Phi \star q_{\rm F}(\rr,\alpha) , \\ 
q_{\rm B}(\rr, \alpha-1) = \e^{- W_{\alpha-1}(\rr)} \, 
\Phi \star q_{\rm B}(\rr,\alpha) ,
\end{aligned}
\end{equation}
where $\Phi(\rr) = \e^{-3 \rr^2/2 b^2} / (2 \pi b^2 / 3)^{3/2}$, and 
initiating from $q_{\rm F}(\rr,1) = \e^{- W_{1}(\rr)}$ and 
$q_{\rm B}(\rr,N) = \e^{-W_N(\rr)}$. Given $q_{\rm F,B}$, 
the polymer bead density operator can be computed as
\begin{equation}
\tilde{\rho}_{\rm b}(\rr) = \Gamma \star \frac{\rho_{\rm p}}{Q_{\rm p}}
\sum_{\alpha = 1}^N \e^{W_{\alpha}(\rr)} q_{\rm F}(\rr,\alpha) 
q_{\rm B}(\rr,\alpha) ,
\end{equation}
with
\begin{equation}
\label{eq:Qp2}
Q_{\rm p} = \frac{1}{V} \int \d \rr \, q_{\rm F}(\rr,N) .
\end{equation}
In the second relation in Eq.~\eqref{eq:H_derivatives}, 
$\tilde{c}(\rr) = \tilde{c}_{\rm b}(\rr) + \tilde{\rho}_+(\rr) 
- \tilde{\rho}_- (\rr)$ is a field operator corresponding to the 
electric charge density, wherein the polymer bead contribution 
$\tilde{c}_{\rm b}(\rr)$ is given by
\begin{equation}
\tilde{c}_{\rm b}(\rr) = \Gamma \star \frac{\rho_{\rm p}}{Q_{\rm p}} \sum_{\alpha = 1}^N \e^{W_{\alpha}(\rr)} q_{\rm F}(\rr,\alpha) q_{\rm B}(\rr,\alpha) \sigma_{\alpha} .
\end{equation}
In the third relation in Eq.~\eqref{eq:H_derivatives},
the field operators for the nonelectric charge densities associated
with the spatially short-range interactions,
$\tilde{h}_a(\rr)$, are similarly given by
\begin{equation}
\label{eq:ha}
\tilde{h}_a(\rr) = \Gamma \star \frac{\rhop}{Q_{\rm p}} \sum_{\alpha = 1}^N
\e^{W_{\alpha}(\rr)} q_{\rm F}(\rr,\alpha) q_{\rm
B}(\rr,\alpha)q^{(a)}{}_{r_\alpha} \; .  
\end{equation}
It can be readily verified that the position-independent solution 
to $\delta H / \delta \phi(\rr)=0$ for $\phi=\eta,\psi,\varphi_a$ 
(where the expressions for the functional derivatives are given in 
Eq.~\eqref{eq:H_derivatives}) 
yields exactly the MFT solution in Eq.~\eqref{eq:MFT_solution} for 
position-independent $\eta$ and $\varphi_a$ (together with any value 
of position-independent $\psi$): For any set of values of the 
position-independent fields, the field operators for number and charge 
densities become equal to their bulk counterparts, 
i.e.~$\tilde{\rho}_{\rm b, \pm, w}(\rr)=\rho_{\rm b, \pm, w}$, 
$\tilde{c}(\rr) = \rho_{\rm p} \sigma_{\rm p} + \rho_+ - \rho_- = 0$ and 
$\tilde{h}_a(\rr) = \rho_{\rm p} q^{(a)}$. Substituting this into 
Eq.~\eqref{eq:H_derivatives} and setting the functional derivatives 
to zero gives exactly the expressions in Eq.~\eqref{eq:MFT_solution}.


In contrast to the approximate analytical approach introduced
under the previous subheading, FTS depends on the full analytical 
form of the spatially short-range interaction potential $V_{\rm h}(r)$,
where $r=|\rr|$, not only its spatial integral
$\int\d\rr V_{\rm h}(r)=\hat{V}_{\rm h}(\bm{0})$ that appears as part of
an effective Flory-Huggins $\chi$-parameter,
$\chi_{\rm eff}$, in Eq.~\eqref{eq:chi_eff}.
Several functional forms for spatially short-range interactions have 
been used extensively in recent coarse-grained simulations of IDP 
LLPS.\cite{dignon18,suman2,Mpipi} These include
the common Lennard-Jones (LJ) potential
\begin{equation}
\label{eq:LJ}
V_{\rm LJ}({\cal E},\sigma | r)
= 4{\cal E}\left [ \left(\frac {\sigma}{r}\right)^{12}
-\left(\frac {\sigma}{r}\right)^{6} \right] \; , 
\end{equation}
where ${\cal E}$ is the depth of the potential well and $\sigma$ is
a length scale, and LJ variations such as the shifted Weeks–Chandler–Andersen 
(WCA) potential\cite{WCA} for finite-range purely repulsive interactions
as well as the recently proposed Wang-Frenkel potential\cite{WangFrenkel} 
with a finite spatial range $r_{\rm c}$,
\begin{equation}
\label{eq:WF1}
V_{\rm WF}({\cal E},\mu,\nu,r_{\rm c},\sigma | r)=
{\cal E}\alpha_{\rm WF}(\mu,\nu,r_{\rm c},\sigma)
\left[\left(\frac {\sigma}{r}\right)^{2\mu}-1\right] 
\left[\left(\frac {r_{\rm c}}{r}\right)^{2\mu}-1\right]^{2\nu} 
\; 
\end{equation}
for $r\leq r_{\rm c}$ and $V_{\rm WF}(\mu,\nu,r_{\rm c},\sigma | r)=0$ for
$r>r_{\rm c}$ with
\begin{equation}
\label{eq:WF2}
\alpha_{\rm WF}(\mu,\nu,r_{\rm c},\sigma) =
2\nu\left(\frac {r_{\rm c}}{\sigma}\right)^{2\mu}
\left\{\frac {1+2\nu}{2\nu[(r_{\rm c}/\sigma)^{2\mu}-1]}\right\}^{2\nu+1}
\; ,
\end{equation}
that enjoys several apparent numerical advantages when applied
to many-body simulations, including its smooth decay to zero as
$r$ is increased toward the cutoff distance $r_{\rm c}$.
Note that the original notation in ref.~\citen{WangFrenkel} for
the Wang-Frenkel form is largely
followed in Eqs.~\eqref{eq:WF1} and \eqref{eq:WF2}. The symbols 
$\alpha_{\rm WF}$ and $\mu$ here should not be confused with the polymer 
bead label $\alpha=1,2,\dots,N$ and the symbol for chemical potential.
The length scale $\sigma$ in Eqs.~\eqref{eq:LJ}--\eqref{eq:WF2}
also should not be confused with the symbol for electric charge
defined above in our formulation.

Mathematically, however, these potential functions cannot practically---if
at all possibly---be incorporated into our field-theoretic formalism because
the field Hamiltonian $H$ in Eq.~\eqref{eq:field_hamiltonian} requires 
the inverse operator, ${V_{\rm h}}^{-1}$, of the potential function 
but no inverse operator expressed in terms of elementary functions
is known for $V_{\rm LJ}$ and $V_{\rm WF}$. 
(For a recent promising approach for including general pairwise interaction
potentials in FTS, see ref.~\citen{ottinger2021}).
This consideration leads us to
the Yukawa potential,\cite{yukawa} which has 
the same mathematical form as the screened Coulomb potential,  
because it possesses an inverse expressible in terms of Fourier transform
of elementary functions. Therefore, as a first step in the exploration
of our general theoretical framework, we specialize here to a Yukawa form
for the spatially short-range interactions, viz.,
\begin{equation}
\label{eq:yukawa}
V_{\rm h}(r) = \frac{l_{\rm h} \e^{-\kappa r}}{r} \, ,
\end{equation}
where interaction strength is represented by $l_{\rm h}$ and the
spatial range of the interaction is characterized by the reciprocal
$\kappa^{-1}$ of the screening coefficient $\kappa$ (Fig.~4). 
The Fourier transform of $V_{\rm h}(r)$ is given by
$\hat{V}_{\rm h}(\kk) = \hat{V}_{\rm h}(k) = 
4 \pi l_{\rm h} /  (k^2 + \kappa^2)$, and
the inverse of $V_{\rm h}(r)$ as defined by Eq.~\eqref{eq:Vh} is 
${V_{\rm h}}^{-1} = (-\bm{\nabla}^2 + \kappa^2)/4 \pi l_{\rm h}$.
In view of the MFT result in Eq.~\eqref{eq:chi_eff}, the overall 
spatial range-dependent interaction strength of the Yukawa potential may 
also be quantified by $\hat{V}_{\rm h} (\bm{0}) = 4 \pi l_{\rm h} / \kappa^2$. 
\\

\begin{figure}[!t]
\centering
   \includegraphics[width=0.45\columnwidth]{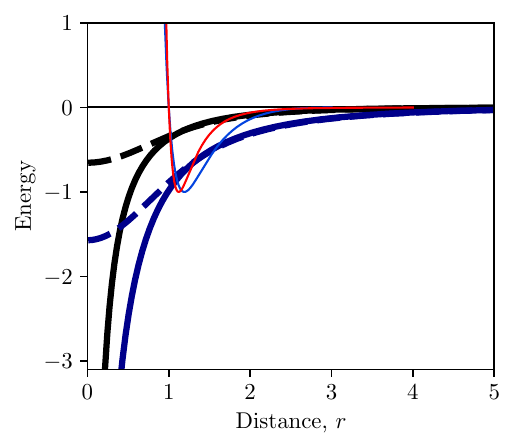}
\vskip -2 mm
\caption*{\footnotesize {\bf Fig.~4:} 
The Yukawa potential as a model of spatially short-range favorable 
interactions. Two examples of $V_{\rm h}(r)$ 
in Eq.~\eqref{eq:yukawa} are depicted by
the thick solid black curve ($l_{\rm h}=6\sigma/3.8$, $\kappa=3.8/8\sigma$)
and the thick solid blue curve ($l_{\rm h}=3\sigma/3.8$, 
$\kappa=3.8/5\sigma$) with $\sigma=1$; i.e., $\sigma$ is the unit for $r$
in this plot.
The dashed curves (same color code) provide the corresponding ``smeared'' 
version of these potentials defined as
$(1/2\pi^2)\int_0^\infty \d k k^2 \hat{\Gamma}(k)^2 \hat{V}(k)
\sin(kr)/kr$ where $\hat{\Gamma}(k)=\exp(-a_{\rm s}^2k^2/2)$ is the Fourier
transform of the function $\Gamma(\rr)$
in Eq.~\eqref{eq:Gamma} with $a_{\rm s}=\sigma/\sqrt{6}$ shown 
as an example here.
The interaction between two point particles via this smeared Yukawa potential
is mathematically equivalent to that between two particles modeled
as Gaussian distributions $\Gamma(\rr)$ via the original $V_{\rm h}(r)$.
Included for comparison are a LJ potential
(red curve, Eq.~\eqref{eq:LJ} with ${\cal E}=1$)
and a Wang-Frenkel potential (thin blue curve, Eq.~\eqref{eq:WF1} with
${\cal E}=1$, $\mu=\nu=1$, and $r_{\rm c}=3\sigma$).
The example Yukawa potentials are selected to match
approximately with either the potential wells 
(thick blue curves) or the large-$r$ trends (thick black curves) 
of the LJ and Wang-Frenkel potentials.
}
\label{ref:fig4}
\end{figure}

{\bf Pair Correlation Functions.}
Structural information of the molecular species in our model system
can be inferred from pair correlation functions,\cite{Pal2021,wessen2022}
defined here as
\begin{equation} \label{eq:corr_func_def}
G_{p,q}(|\rr-\rr'|) = \langle \hat{\rho}_p(\rr) \hat{\rho}_q(\rr') \rangle -
\delta_{pq} \rho_q 
\frac{\e^{-|\rr-\rr'|^2/4 a_{\rm s}^2}}{(4 \pi a_{\rm s}^2)^{3/2}} \; ,
\quad \quad p,q = {\rm b, \pm, w} \; ,
\end{equation}
where the second term on the right hand side of the equation for $G_{p,q}$ 
subtracts the trivial correlation of a particle with itself. 
Correlation functions were used to quantify the degree of
condensed-state subcompartmentalization
of phase-separated polyampholytes in FTS and coarse-grained explicit-chain
models\cite{Pal2021} and have been applied more recently to analyze
experimental data on the biomolecular structure
of nucleolar condensates.\cite{JoshCliff}
The pair correlation functions are, in turn, related to the potentials 
of mean force (PMFs), $U_{p,q}(r)$, through the standard formula
\begin{equation} \label{eq:pmf_def}
U_{p,q}(r) = - k_{\rm B} T \ln\left[\frac{G_{p,q}(r)}{\rho_p \rho_q}\right] 
\, , 
\end{equation}
where the normalization factor $\rho_p \rho_q$ in the argument of
the logarithm is chosen such that $U_{p,q}<0$ means
correlation and $U_{p,q}>0$ implies anti-correlation. In FTS, the pair 
correlation functions can be computed using the following relations:
\begin{equation}
\begin{aligned}
\langle \hat{\rho}_p(\rr) \hat{\rho}_q(\rr') \rangle &= \langle
\tilde{\rho}_p(\rr) \tilde{\rho}_q(\rr') \rangle \quad , \quad (p \neq q) , \\
\langle \hat{\rho}_p(\rr) \hat{\rho}_p(\rr') \rangle &= \left\langle
\frac{\tilde{\rho}_p(\rr)}{v_p} \left( \frac{\i}{\gamma} \eta(\rr') + \rho_0 -
\sum_{q\neq p} v_q \tilde{\rho}_q(\rr') \right) \right\rangle \; .
\end{aligned}
\end{equation}
In accordance with a common procedure in field theory,\cite{IZ}
these relations can be derived by adding source terms $\sum_p \int \d \rr
\hat{\rho}_p(\rr) J_p(\rr)$ to the particle Hamiltonian $\hat{H}$ in
Eq.~\eqref{eq:particle_H} and then taking functional derivatives 
with respect to $J_p(\rr)$s in the resulting field theory and finally
setting all $J_p \rightarrow 0$. The
formula for the self-correlation function $\langle \hat{\rho}_p \hat{\rho}_p
\rangle$ may be obtained\cite{Pal2021} through an intermediate 
field re-definition to avoid a term that contains a double 
derivative of $Q_p$. 
\\


{\bf Coarse-Grained Explicit-Chain Models.}
Following our previous studies,\cite{SumanPNAS,Pal2021,wessen2021}
coarse-grained explicit-chain molecular dynamics (MD) simulations
are conducted here to elucidate the ramifications of several
proposed interaction schemes\cite{dignon18,SumanPNAS,Urry-Mittal,FB,KH,Mpipi} 
and to assess the accuracy of our approximate analytical
theories and FTS in capturing the essential physics of biomolecular LLPS.
As before,\cite{dignon18,suman2}
each amino acid is represented by a single bead of different mass and
size. Although it is possible to incorporate explicit simple dipole solvent
molecules in coarse-grained explicit-chain LLPS simulation to better
account for dielectric effects,\cite{wessen2021}
for simplicity we will limit the present study to implicit-solvent
simulations.\cite{dignon18,SumanPNAS}
As the general features of the model can be found in the
literature,\cite{dignon18,SumanPNAS,Urry-Mittal,FB,KH,Mpipi}
only an outline is provided here. Briefly, as in the above field-theory
formulation, let $n_{\rm p}$ be the total number of $N$-residue polypeptide 
chains (IDPs) in the MD system, with the chains labeled by $i$ 
or $j=1,2,\dots,n_{\rm p}$ and the residues (beads) along a polypeptide
chain labeled by $\alpha$ or $\beta=1,2,\dots,N$.
The total potential energy $U_{\rm T}$ is a function of
the residue positions $\{\RR_{i,\alpha}\}$ that may be written as
\begin{equation}
U_{\rm T} = U_{\rm bond} + U_{\rm el} + U_{\rm hh} \; ,
\label{eq:U-T}
\end{equation}
where $U_{\rm bond}$ is the bond-length term for chain connectivity:
\begin{equation}
U_{\rm bond} = \frac {K_{\rm bond}}{2} \sum_{i=1}^{n_{\rm p}} 
\sum_{\alpha=1}^{N-1}
(R_{i,\alpha+1;i,\alpha}-l)^2
\;
\label{eq:U-bond}
\end{equation}
with $R_{i,\alpha;j,\beta}\equiv |\RR_{i,\alpha}-\RR_{j,\beta}|$.
Parameter values for the MD potential $U_{\rm bond}$ in Eq.~\eqref{eq:U-bond} 
and for $U_{\rm el}$, $U_{\rm hh}$ below are different 
across the three interaction schemes---Mpipi,\cite{Mpipi}
Urry,\cite{Urry-Mittal} and FB\cite{FB}---that we apply here to the 
Ddx4 IDRs.\cite{SumanPNAS} 
To facilitate comparisons with prior coarse-grained MD simulation
results in the literature, the parameters in these original
references are largely used with the corresponding interaction schemes
in the present simulations unless stated otherwise. For $U_{\rm bond}$,
$K_{\rm bond}=19.1917$ kcal mol$^{-1}$\AA$^{-2}$, $l=3.81$ \AA~for
Mpipi;\cite{Mpipi} 
$K_{\rm bond}=9.56$ kcal mol$^{-1}$\AA$^{-2}$, $l=3.82$ \AA~for
Urry;\cite{Urry-Mittal}
and
$K_{\rm bond}=2.39$ kcal mol$^{-1}$\AA$^{-2}$, $l=3.8$ \AA~for
FB.\cite{FB}
The $U_{\rm el}$ term in Eq.~\eqref{eq:U-T} is the electrostatic interaction:
\begin{equation}
U_{\rm el} =
\sum_{i,j=1}^{n_{\rm p}}
\sum_{\substack{\alpha,\beta=1\\ \null\hskip -0.8cm(i,\alpha)\ne(j,\beta)}}^N
\frac {\sigma_{i,\alpha}\sigma_{j,\beta}e^2}{4\pi\epsilon_0\epsilon_{\rm r}
R_{i,\alpha;j,\beta}} \exp \Bigr(-\kappa_{\rm D} R_{i,\alpha;j,\beta} \Bigr)
\; ,
\label{eq:U-el}
\end{equation}
where, as in Eq.~\eqref{eq:hatc}, $\sigma_{i,\alpha}$ is the electric charge,
in units of proton charge $e$, of the $\alpha$th residue along the $i$th chain 
(as stated above for the field-theoretic formulation, 
$\sigma_{i,\alpha}=\sigma_\alpha$ is independent of $i$ in
the present MD simulations), and $\kappa_{\rm D}$ is the reciprocal
of the Debye screening length. 
Mpipi\cite{Mpipi} uses $\sigma_\alpha=+0.75$ for the positively charged
residues arginine and lysine, 
$\sigma_\alpha=+0.375$ for histidine (to account for its incomplete ionization
under physiological conditions), and
$\sigma_\alpha=-0.75$
for the negatively charged residues
aspartic and glutamic acids;
Urry\cite{Urry-Mittal} uses $\sigma_\alpha=+1$ for arginine and lysine,
$\sigma_\alpha=0$ for histidine, and $\sigma_\alpha=-1$ for aspartic
and glutamic acids;
and FB\cite{FB} uses $\sigma_\alpha=+1$ for arginine and lysine, 
$\sigma_\alpha=+0.5$ for histidine, and $\sigma_\alpha=-1$ for aspartic
and glutamic acids. All other residues have $\sigma_\alpha=0$ in 
all of these three interaction schemes.
In all of the present coarse-grained MD simulations, we set
$\kappa_{\rm D}=0.1$ \AA$^{-1}$ (i.e., a Debye screening length
of 10 \AA~corresponding to a physiological monovalent
salt concentration of $\sim 100$ mM), apply a large-distance 
cutoff of 35 \AA~to the electrostatic interaction,
and use a dielectric constant $\epsilon_{\rm r}=40$. Employing
a dielectric constant moderately lower than the 
$\epsilon_{\rm r}\approx 80$ value of bulk water
in coarse-grained MD simulation of IDP LLPS has recently been
rationalized from a theoretical perspective\cite{SumanPNAS,wessen2021} and
this modeling practice is apparently not inconsistent with experimental 
observations of slower water orientational dynamics in 
living matter.\cite{Tros2017}

For the spatially short-range interaction term $U_{\rm hh}$
in Eq.~\eqref{eq:U-T}, Mpipi\cite{Mpipi} uses the Wang-Frenkel 
form,\cite{WangFrenkel} viz.,
\begin{equation}
\label{eq:UhhMpipi}
U_{\rm hh}^{\rm Mpipi}=
\sum_{i,j=1}^{n_{\rm p}}
\sum_{\substack{\alpha,\beta=1\\ \null\hskip -0.8cm(i,\alpha)\ne(j,\beta)}}^N
V_{\rm WF}({\cal E}_{\alpha,\beta},\mu_{\alpha,\beta},1,(r_{\rm c})_{\alpha,\beta},\sigma_{\alpha,\beta} 
| R_{i,\alpha;j,\beta}) \, ,
\end{equation}
where, by definition, $V_{\rm WF}({\cal E}_{\alpha,\beta},\mu_{\alpha,\beta},1,(r_{\rm c})_{\alpha,\beta},\sigma_{\alpha,\beta}| R_{i,\alpha;j,\beta})$
vanishes for $R_{i,\alpha;j,\beta}> (r_{\rm c})_{\alpha,\beta}$ as specified
in Eqs.~\eqref{eq:WF1} and~\eqref{eq:WF2}. For all amino acid residue 
pairs $\alpha,\beta$, Mpipi sets the distance cutoff
$(r_{\rm c})_{\alpha,\beta}=3\sigma_{\alpha,\beta}$ and assigns $\nu=1$.
Values of the interaction strength ${\cal E}_{\alpha,\beta}$, 
length scale $\sigma_{\alpha,\beta}$, and the parameter
$\mu_{\alpha,\beta}$ for the Wang-Frenkel functional form 
depend on the amino acid residue types
$r_\alpha$ and $r_\beta$ for residues $\alpha$ and $\beta$.
The Mpipi values for ${\cal E}_{r,r'}$, $\sigma_{r,r'}$, and
$\mu_{r,r'}$, where $r$ and $r'$ each represents one of the twenty
amino acid types, 
are provided in Supplementary Table~11 of ref.~\citen{Mpipi}, wherein
$\mu_{r,r'}=2$ for almost all pairs of residue types except for
a few cases in which $\mu_{r,r'}=4$ or $11$.

For the Urry\cite{Urry-Mittal} and FB\cite{FB} interaction schemes, 
$U_{\rm hh}$ in Eq.~\eqref{eq:U-T} is given by\cite{dignon18}
\begin{equation}
\label{eq:Uhh0}
U_{\rm hh}^{\rm Urry/FB}=
\sum_{i,j=1}^{n_{\rm p}}
\sum_{\substack{\alpha,\beta=1\\ \null\hskip -0.8cm(i,\alpha)\ne(j,\beta)}}^N
(U_{\rm hh})_{i,\alpha;j,\beta}^{\rm Urry/FB}
\end{equation}
where
\begin{equation}
\label{eq:Uhh}
(U_{\rm hh})_{i,\alpha;j,\beta}^{\rm Urry/FB} =
\begin{cases}
V_{\rm LJ}({\cal E}_{\alpha,\beta},\sigma_{\alpha,\beta}|R_{i,\alpha;j,\beta})
 + (1-\lambda_{\alpha,\beta}^{\rm hh}){\cal E}_{\alpha,\beta}
& {\rm if\ } R_{i,\alpha;j,\beta}\le 2^{1/6}\sigma_{\alpha,\beta}\\
\lambda_{\alpha,\beta}^{\rm hh} 
V_{\rm LJ}({\cal E}_{\alpha,\beta},\sigma_{\alpha,\beta}|R_{i,\alpha;j,\beta})
& {\rm otherwise}
\end{cases}
\;
\end{equation}
is in the Ashbaugh-Hatch form\cite{AH}
with the function $V_{\rm LJ}$ given by Eq.~\eqref{eq:LJ}.
For both Urry and FB, ${\cal E}_{\alpha,\beta}={\cal E}=0.2$ kcal mol$^{-1}$
irrespective of the residue types $r_\alpha,r_\beta$ of 
residues $\alpha,\beta$, and the length scale
$\sigma_{\alpha,\beta}=[\sigma(r_\alpha)+\sigma(r_\beta)]/2$, where
$\sigma(r_\alpha)$ is the diameter of the bead model for amino acid
residue type $r_\alpha$ given by the quantity
$\sigma$ in Table~S1 of ref.~\citen{dignon18}. 
In both interaction schemes, 
$\lambda_{\alpha,\beta}^{\rm hh}=[\lambda(r_\alpha)+\lambda(r_\beta)]/2$,
where $\lambda(r_\alpha)$ is the hydrophobicity/hydropathy parameter
for residue type $r_\alpha$. The $\lambda(r_\alpha)$ values for the
Urry interaction scheme are given by the ``Urry et al. normalized
hydropathy scale'' column of Table~S2 in ref.~\citen{Urry-Mittal}, 
whereas the corresponding $\lambda(r_\alpha)$ values for the FB interaction
scheme are provided in Table~S7 of ref.~\citen{FB}.
In simulations that use the Urry and FB interaction schemes, a cutoff 
distance of
20 \AA~is applied to the $(U_{\rm hh})_{i,\alpha;j,\beta}^{\rm Urry/FB}$
interactions in Eq.~\eqref{eq:Uhh}.

In addition to modeling Ddx4 IDRs, 
to investigate nonelectrostatic effects of sequence pattern on LLPS,
coarse-grained explicit-chain MD 
simulations are also conducted here for copolymer sequences consisting of 
only two types of residues, both with zero electric charge, 
but possess different hydrophobicities.
The hydrophobic-polar patterns of these model sequences---commonly referred 
to as HP sequences in the protein 
literature\cite{irback2020,dill1985,lau1989,panag1992}---are taken 
from recent studies of their LLPS.\cite{Statt2020,panagio2021}
In keeping with tradition of surfactant models, however, hydrophobic 
and polar beads were labeled, respectively, as ``T'' (tail) and 
``H'' (head) in refs.~\citen{Statt2020,panagio2021}.
Here we use leucine for the hydrophobic beads and serine for the polar
beads. The resulting amino acid sequences are thus referred to as 
LS sequences. As a test case, we use the KH interaction scheme\cite{dignon18}
described in ref.~\citen{SumanPNAS} for the present LS-sequence simulations.
Specifically, the spatially short-range interaction $U_{\rm hh}$ is
in the form of Eqs.~\eqref{eq:Uhh0} and \eqref{eq:Uhh}
wherein ${\cal E}_{\alpha,\beta}=0.228|(e_{\rm MJ})_{r_\alpha,r_\beta}-e_0|$,
$\lambda^{\rm hh}_{\alpha,\beta}=1$ if 
$(e_{\rm MJ})_{r_\alpha,r_\beta}\leq e_0$ and 
$\lambda^{\rm hh}_{\alpha,\beta}=-1$ otherwise,
$e_0=-1.0$ kcal mol$^{-1}$ and 
$(e_{\rm MJ})_{r_\alpha,r_\beta}$ is from ref.~\citen{MJ96}.
The KH-D parameter set in Table~S3 of ref.~\citen{dignon18}
corresponds to
$-\lambda^{\rm hh}_{\alpha,\beta}{\cal E}_{r_\alpha,r_\beta}$.
\\

{\bf MD Simulation of LLPS.}
Based on the models described above,
we follow the recently developed, widely-applied slab 
method for simulation of IDP LLPS.\cite{dignon18,suman2,mimb2022}
Because this simulation protocol, its rationale, and the procedure to 
construct phase diagrams from equilibrated simulations have been
detailed elsewhere,\cite{dignon18,suman2,panag2017}
a brief outline here will suffice. As in our previous 
studies,\cite{suman2,Pal2021,SumanPNAS,mimb2022,wessen2021}
the GPU version of HOOMD-Blue software\cite{HOOMD,Anderson}
is employed in the present simulations.  For each simulation of the wildtype or
a variant of the 236-residue Ddx4 IDR,\cite{SumanPNAS}
100 chains are randomly placed in a sufficiently large cubic
box of dimension $300\times 300\times 300$ \AA$^3$ initially. 
Energy minimization is
then performed to remove any steric clashes among the model molecules. 
This is followed by $NPT$ compression at a temperature of 100 K and
pressure of 1 atm for a period of 50 ns using the Martyna-Tobias-Klein
(MTK) thermostat and barostat\cite{klein1994,martyna2006} 
with a coupling constant of 1 ps. The equations of
motion are integrated with velocity-Verlet algorithm using a 
large timestep of 10 fs. Periodic boundary conditions are applied in 
all three directions. The electrostatic interaction is
treated with the PPPM algorithm.\cite{LeBard} 
After the initial $NPT$ compression, the system is compressed again
at 100 K for a period of 50 ns using Langevin thermostat with 
a friction coefficient of 1 ps$^{-1}$ to reach a sufficiently high density 
phase of IDPs enclosing in a simulation box of size
$150\times 150\times 150$ \AA$^3$. At this point,
the system is expanded along one of the axes (referred to as
the $z$-axis) to 3000 \AA~for 10 ns with the temperature fixed at 100 K. 
Next, an $NVT$ equilibration is performed for 2 $\mu$s at desired 
temperatures using the Langevin thermostat with a friction coefficient 
of 1 ps$^{-1}$. Final production run is then carried out for another 4 $\mu$s 
with the same Langevin thermostat using a lower friction coefficient 
of 0.01 ps$^{-1}$. Snapshots are saved every 1 ns for analysis. 
A similar procedure is used for the simulations of the 
20-residue leucine-serine sequences LS1, LS2, and LS3.
Here, the only difference with the Ddx4 IDR simulations is that 
1,000 LS chains are used for each simulation and 
the simulation box is compressed to a size of
$130\times 130\times 130$ \AA$^3$ and then
expanded to 2000 \AA~along the $z$-axis.
Details regarding how to construct phase diagrams
from equilibrium trajectories are described in 
refs.~\citen{dignon18,suman2,SumanPNAS,mimb2022}.
\\

$\null$

\noindent
{\large\bf RESULTS AND DISCUSSION}\\

While we consider the above-detailed theoretical development to be
the main thrust of this article and that extensive testing of our theory
is beyond the scope of the present work, it is instructive
to apply the new formulation to a few initial examples to probe our 
theory's practical effectiveness and to provide suggestions for future 
theoretical/computational improvements based upon the success and 
limitation of the following examinations. 
We do so by using our formulation on four well-studied 
Ddx4 IDRs\cite{Nott15,jacob2017,SumanPNAS,robert,Urry-Mittal,Mpipi}
as well as three selected hydrophobic-polar model sequences that have
the same composition but different hydrophobic/polar sequence 
patterns.\cite{Statt2020}
\\


{\bf Analytical Theory for the Phase Behaviors of Wildtype and Variant 
Ddx4 IDRs Modeled by Different Interaction Schemes.} 
To illustrate the approximate analytical approach in the previous
section, we apply the theory with the free energy function given
by Eq.~\eqref{eq:MFT_free_energy} to the 236-residue wildtype (WT),
charge-scrambled (CS), arginine-to-lysine (RtoK), and
phenylalanine-to-alanine (FtoA) variants of the Ddx4 IDR (the amino 
acid sequences are provided, for example, in Fig.~S1 of
ref.~\citen{SumanPNAS}). We compare the LLPS properties predicted 
by using $\varepsilon_{r,r'}$s 
from the five interaction schemes illustrated in Fig.~2.
In accordance with Eqs.~\eqref{eq:chi_eff} and \eqref{eq:lambda-sum} for 
$\chi_{\rm eff}$, all nonzero eigenvalues $\lambda_a$s are included in 
the application of the MFT/RPA Eq.~\eqref{eq:MFT_free_energy} to
the four Ddx4 IDRs.

\begin{figure}[!ht]
\centering
   \includegraphics[width=0.75\columnwidth]{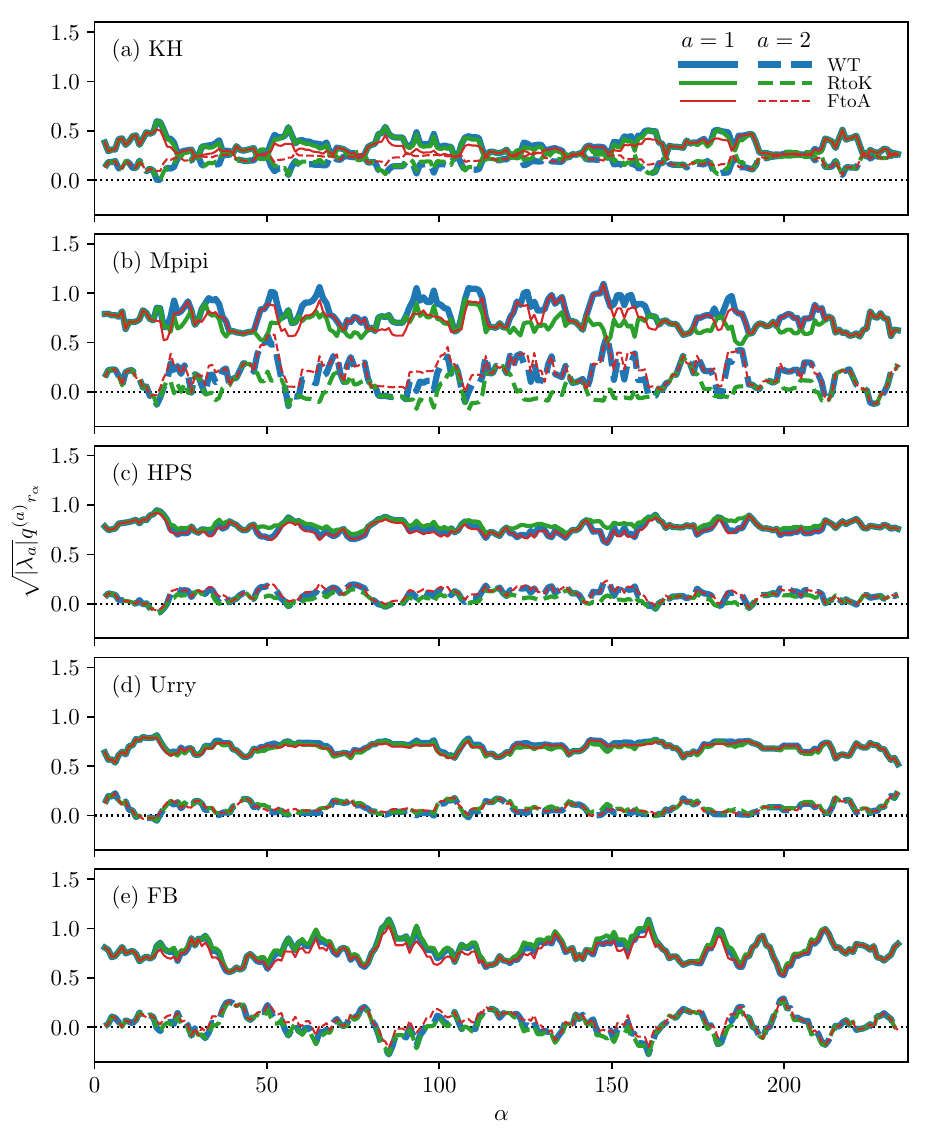}
\vskip -2 mm
\caption*{\footnotesize {\bf Fig.~5:} 
Nonelectric charge profiles for Ddx4 IDRs.
Distributions of nonelectric charges $q^{(a)}{}_{r_\alpha}$
for the two most dominant eigenvalues $a=1,2$ of $\varepsilon_{r,r'}$
[Eq.~\eqref{eq:eps_spectral_decomposition}] 
along the WT, RtoK and FtoA Ddx4 sequences are shown
in the (a) KH, (b) Mpipi, (c) HPS, (d) Urry, and (e) FB interaction schemes
[key for line styles provided in (a)].
The nonelectric charge profiles in this figure are 
averages of individual residues' $q^{(1)}{}_{r_\alpha}$ or 
$q^{(2)}{}_{r_\alpha}$ over a sliding window of six residues.
As generally defined in the text, $\alpha=1,2,\dots,236$ is the 
sequential position label of the residues along the Ddx4 IDRs, 
and $r_\alpha$ is the amino acid type of residue $\alpha$.
}
\label{ref:fig5}
\end{figure}

Fig.~5 shows the nonelectric charge profiles of the WT, RtoK and FtoA 
Ddx4 IDRs for the two most dominant eigenvalues $\lambda_{1,2}$ in each
of the five interaction schemes considered. 
Significant differences in nonelectric
charge profiles are observed across different interaction schemes
(Fig.~5a--e), reflecting variations in the assumptions 
made by KH, Mpipi, HPS, Urry, and FB regarding physical driving
forces in biomolecular processes.
Variation in the two dominant nonelectric charges 
along the IDR sequences (as function of the horizontal
varible $\alpha$) is discernibly more pronounced in Mpipi and FB 
than in HPS and Urry, with KH exhibiting an intermediate degree for
variation along the chain sequence. Implications of this difference
for predicted LLPS propensity remain to be ascertained.
Within the same interaction scheme, the variation in nonelectric
charge profiles among the three Ddx4 sequence variants (shown by
curves in different colors in Fig.~5) is notably higher
for the KH and Mpipi interaction schemes with 210 contact energies
than for the HPS, Urry, and FB interaction schemes that are based upon
20-value hydrophobicity/hydropathy scales, indicating that the
LLPS propensities of the sequence variants predicted by KH and Mpipi
may be more diverse in KH and Mpipi than those predicted
by HPS, Urry, and FB.

Because the WT, RtoK and FtoA Ddx4 IDRs share the same sequence electric charge 
pattern, any sequence dependence of LLPS propensity
among these three sequences in the MFT/RPA analytical 
approach originates from the difference in their 
$\chi_{\rm eff}$ parameters, which may be computed 
from the eigenvalues $\lambda_a$ of the spectral decomposition
of $\varepsilon_{r,r'}$
in conjunction with the nonelectric charges $q^{(a)}{}_{r_{\alpha}}$ using
Eq.~\eqref{eq:chi_eff} or directly from the interaction matrix 
$\varepsilon_{r,r'}$ itself via 
the identity in Eq.~\eqref{eq:lambda-sum}. In the present MFT/RPA formulation,
which for simplicity assigns the same $V_{\rm h}(r)$ to every residue-residue 
interaction scheme considered, 
the ratios between these $\chi_{\rm eff}$ parameters 
depend only on $\varepsilon_{r,r'}$ but not $V_{\rm h}(r)$. 
As such, these ratios may serve as a zeroth order
approximate measure of the relative contributions to LLPS propensity from 
the spatially short-range, nonelectrostatic interactions encoded by 
the different IDR sequences.
Accordingly, for the five interaction schemes we consider, 
the relative short-spatial-range interaction strengths among
the WT, CS, RtoK and FtoA Ddx4 IDRs is quantified by MFT in terms of
the ratios
\begin{equation}
\chi^{\rm (WT,CS)}_{\rm eff} : \chi^{\rm (RtoK)}_{\rm eff} : 
\chi^{\rm (FtoA)}_{\rm eff} 
\; 
\end{equation}
wherein WT and CS share the same $\chi_{\rm eff}$
because the WT and CS sequences have the same amino acid composition.
Utilizing the $\varepsilon_{r,r'}$ values of the interaction
schemes\cite{dignon18,Urry-Mittal,FB,Mpipi}
in Eqs.~\eqref{eq:chi_eff} and \eqref{eq:lambda-sum}
leads to the following ratio:
\begin{equation}
\label{eq:chi-ratios}
\begin{aligned}
&\chi^{\rm (WT,CS)}_{\rm eff} : \chi^{\rm (RtoK)}_{\rm eff} : \;
\chi^{\rm (FtoA)}_{\rm eff} &  \\
&\quad\quad\quad 1 \, : \;\; 0.83 \;\, : \;\, 0.66 \quad &(\mbox{KH}) \\
&\quad\quad\quad 1 \, : \;\; 0.80 \;\, : \;\, 0.84 \quad &(\mbox{Mpipi}) \\
&\quad\quad\quad 1 \, : \;\; 1.09 \;\, : \;\, 0.97 \quad &(\mbox{HPS}) \\ 
&\quad\quad\quad 1 \, : \;\; 0.96 \;\, : \;\,  0.97 \quad &(\mbox{Urry}) \\
&\quad\quad\quad 1 \, : \;\; 1.04 \;\, : \;\,  0.94 \quad &(\mbox{FB}) 
\; .
\end{aligned} 
\end{equation}

\begin{figure}[!t]
\centering
   \includegraphics[width=0.95\columnwidth]{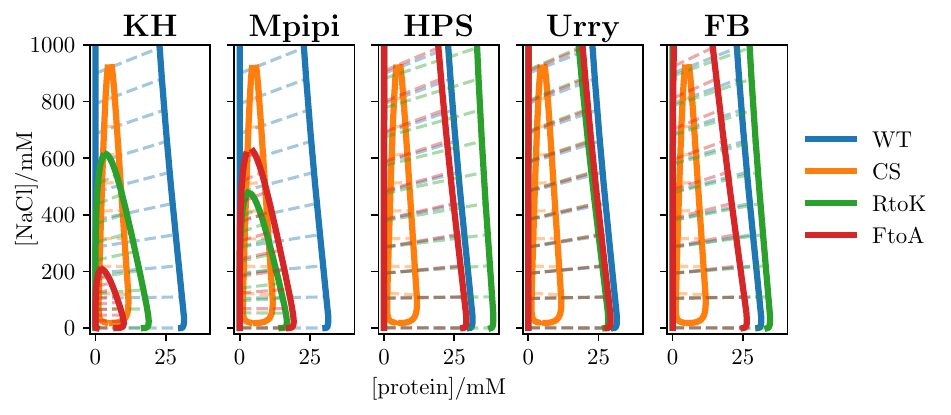}
\vskip -2 mm
\caption*{\footnotesize {\bf Fig.~6:} 
Salt-dependent phase behaviors of Ddx4 IDRs computed by analytical
theory. Phase diagrams of the four Ddx4 variants (color-coded on the right) 
on the protein-salt density $(\rhop,\rho_{\rm s})$ plane are computed
in our approximate MFT/RPA analytical approach 
[Eq.~\eqref{eq:MFT_free_energy}] at model temperature $T=280$ K
using five different interaction schemes as indicated by the top labels.
The phase-separated regimes are enclosed by coexistence (binodal) curves 
(solid curves), $(\rhop,\rho_{\rm s})$ of coexisting phases are connected 
by tie-lines (dashed lines with the same color code for the sequence variants). 
All parameter values used for this figure are described in the text.
}
\label{ref:fig6}
\end{figure}

Phase diagrams can readily be constructed\cite{mimb2022} using the MFT/RPA 
free energy function in Eq.~\eqref{eq:MFT_free_energy}. To facilitate 
comparison of the five different interaction schemes,  the overall
strength of the spatially short-range interactions, encapsulated by 
$\hat{V}_{\rm h}(\bm{0})$, is adjusted for each interaction scheme
separately such that $\chi^{\rm (WT,CS)}_{\rm eff} = 0.5$ at a reference 
temperature of $T=300$ K. We also set $\lB=7$ {\AA} at the same reference 
temperature throughout, and assume the temperature dependence of
$\lB,\hat{V}_{\rm h}(\bm{0}) \propto T^{-1}$ (as stated above, $V_{\rm h}(r)$
is potential energy in units of $k_{\rm B}T$). Other parameters used
are $v_{\rm w}=1$, $v_{\rm b}=2$, $v_{\pm}=0$, $a_{\rm s}=b/\sqrt{6}$, 
and $\rho_0 = 55.5$ M.
In our MFT/RPA calculations for all the interaction schemes considered here,
electric charge is $+e$ for residues R and K, $-e$ for D and E, and zero 
for all the other residues.

Fig.~6 shows phase diagrams in the $(\rhop,\rho_{\rm s})$ plane, where the
two-phase region is enclosed by the coexisting (binodal) curves (solid
lines). Systems with bulk concentrations within the two-phase region
separate into two phases with the partition of molecular species
provided by the tie-lines (dashed lines in Fig.~6).
The phase diagrams in Fig.~6 are computed at a uniform temperature
of $T=280$ K, which is chosen to be sufficiently
low such that every Ddx4 IDR phase separates under any one of the spatially
short-range interaction scheme we consider.

\begin{figure}[!t]
\centering
   \includegraphics[width=0.95\columnwidth]{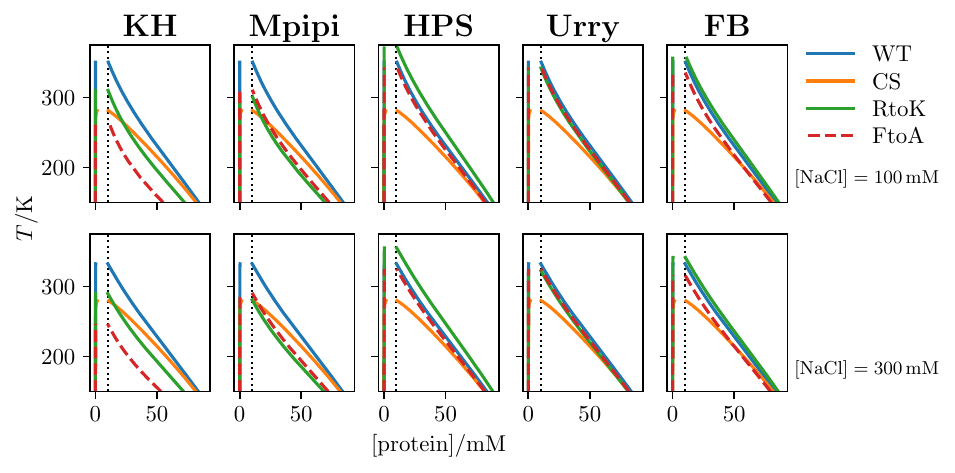}
\vskip -2 mm
\caption*{\footnotesize {\bf Fig.~7:} 
Temperature-dependent phase behaviors of Ddx4 IDRs computed by analytical
theory. As in Fig.~6, phase diagrams are computed in our approximate 
MFT/RPA analytical approach [Eq.~\eqref{eq:MFT_free_energy}]
using five different interaction schemes. Shown here are
protein density ($\rho_{\rm p}$) versus model temperature ($T$)
phase diagrams for overall (bulk) monovalent salt ([NaCl])
concentrations of 100 mM (top row) and 300 mM (bottom row).
These are constructed using constant-$T$ $(\rhop,\rho_{\rm s})$ phase 
diagrams (exemplified by Fig.~6) at multiple $T$. For a given $T$, 
the two coexisting $\rho_{\rm p}$s at bulk salt density 
$\rho_{\rm s}=100$ or $300$ mM 
are obtained as the $\rho_{\rm p}$ values at the two ends
of a $(\rhop,\rho_{\rm s})$ coexistence tie-line
that passes through the intersection point of
$\rho_{\rm p}= 10$ mM (indicated by the vertical dotted lines in
the present figure) and $\rho_{\rm s}=100$ mM or 
$\rho_{\rm s}=300$ mM, respectively, in the $\rhop$--$\rho_{\rm s}$ plane.
The color code for the Ddx4 variants are the same as that in Fig.~6.
Dashed curves are used here for FtoA to facilitate visual
discrimination between the near-overlapping coexisting curves for FtoA (red) 
and RtoK (green) in the Urry interaction scheme.
}
\label{ref:fig7}
\end{figure}

Fig.~7 shows $(\rhop,T)$ phase diagrams of the same systems at a 
constant bulk protein density of $\rhop = 10$ mM (indicated by the 
vertical grey dashed lines), and at two bulk salt densities 
$\rho_{\rm s} = 100$ mM (top) and 300 mM (bottom).
The phase diagrams were constructed by first computing constant-temperature
binodal curves (such as Fig.~6) in the range 
$150\mbox{ K}\leq T\leq 510\mbox{ K}$.
At each temperature, we check if the point 
$(\rhop,\rho_{\rm s})$ lies in the two-phase region and then 
identify the tie-line it intersects. The densities of the coexisting 
phases are then given by the two points on the binodal curve connected 
by the tie-line. It should be noted however that, by construction, the 
coexisting protein densities provided by Fig.~7 for a given temperature
and a given bulk salt concentration apply only to systems with bulk 
protein density $\rhop = 10$ mM.
The coexisting protein densities (i.e., the phase diagram) for the
same temperature and bulk salt concentration would be 
different for systems with a different bulk protein density.
In this regard, the meaning of the phase diagrams in Fig.~7 is more
restrictive than phase diagrams for one solute species in which the
coexisting solute densities for a given temperature apply to
all bulk solute densities within the two-phase region.

Results in Figs.~6 and 7 confirm and add to previous observations
that LLPS properties entailed by different
interaction schemes can be significantly
different.\cite{SumanPNAS,Urry-Mittal,Mpipi}
Two-solute-species $(\rhop,\rho_{\rm s})$ phase properties of Ddx4 IDRs
such as those depicted in Fig.~6 have not been considered before. 
A novel feature that emerges in Fig.~6 is that, for the KH and Mpipi 
interaction schemes, the coexistence curve for the CS variant 
(orange curve) intersects the coexistence curves of other Ddx4 
variants (green and red curves), indicating that the rank ordering
of LLPS propensities is predicted by these interaction schemes to
be salt dependent. This is caused by the fact that 
the sequence electric charge pattern of CS is different from that of the other 
Ddx4 variants and therefore the electrostatic screening effects of salt 
on CS phase separation vary in a different manner from that on the other 
Ddx4 variants when salt concentration varies.
The same feature is manifested 
by the crossing between the CS coexistence curves
and those of other variant(s) in the KH and Mpipi panels
in Fig.~7, indicating that the rank ordering of LLPS propensities
of the Ddx4 IDRs is predicted by these two interaction schemes
to be temperature dependent as well.

As noted before,\cite{SumanPNAS,Mpipi}
for KH and Mpipi, all three variants clearly exhibit lower 
LLPS propensities than the WT (the blue curve is farther to the right
than the other curves in the KH and Mpipi panels in Fig.~6, 
and the blue curve for the condensed-phase protein density
is higher than the other curves in the corresponding panels in Fig.~7). 
This trend is consistent with experiments.\cite{Nott15,jacob2017,robert} 
For the RtoK and FtoA variants, the trend is underpinned by
the effective Flory $\chi$ parameter ratios
$\chi^{\rm (RtoK)}_{\rm eff}/\chi^{\rm (WT)}_{\rm eff}=0.83$,
$\chi^{\rm (FtoA)}_{\rm eff}/\chi^{\rm (WT)}_{\rm eff}=0.66$
for KH and
$\chi^{\rm (RtoK)}_{\rm eff}/\chi^{\rm (WT)}_{\rm eff}=0.80$,
$\chi^{\rm (FtoA)}_{\rm eff}/\chi^{\rm (WT)}_{\rm eff}=0.84$
for Mpipi in Eq.~\eqref{eq:chi-ratios}.
In the Urry interaction scheme, the three Ddx4 variants also show lower 
LLPS propensities than that of the WT, but only barely for RtoK and FtoA
($\chi^{\rm (RtoK)}_{\rm eff}/\chi^{\rm (WT)}_{\rm eff}=0.96$,
$\chi^{\rm (FtoA)}_{\rm eff}/\chi^{\rm (WT)}_{\rm eff}=0.97$ for Urry).
In contrast, as pointed out before,\cite{SumanPNAS} results for HPS in Figs.~6
and 7 indicate that the RtoK variant has a higher LLPS propensity than WT,
which is inconsistent with experiments\cite{Nott15,jacob2017,robert}
($\chi^{\rm (RtoK)}_{\rm eff}/\chi^{\rm (WT)}_{\rm eff}=1.09$,
$\chi^{\rm (FtoA)}_{\rm eff}/\chi^{\rm (WT)}_{\rm eff}=0.97$ for HPS).
The results in Figs.~6 and 7 for FB also exhibit
a higher LLPS propensity for RtoK than for WT, though to a lesser degree.
Similar to HPS, this FB prediction 
($\chi^{\rm (RtoK)}_{\rm eff}/\chi^{\rm (WT)}_{\rm eff}=1.04$,
$\chi^{\rm (FtoA)}_{\rm eff}/\chi^{\rm (WT)}_{\rm eff}=0.94$ for FB)
is caused by its exceptionally high hydrophobicity parameter
of $0.47106$ for lysine compared to
that of $0.24025$ for arginine 
(Table~S7 for optimized FB-HPS parameters in ref.~\citen{FB}).

Experimental data on Ddx4 IDRs indicate that
LLPS propensity of WT is higher than that of CS (ref.~\citen{Nott15})
and that LLPS propensity of CS is higher than that of FtoA 
(ref.~\citen{jacob2017}, FtoA corresponds to 14FtoA in this reference).
Because LLPS for FtoA is observed at $\sim 350$ mg mL$^{-1}$ protein 
concentration but not for RtoK up to $400$ mg mL$^{-1}$ protein 
concentration under the conditions given in Appendix 1--Table~4 
of ref.~\citen{robert}, the overall experimental rank ordering
of LLPS propensities is WT $>$ CS $>$ FtoA $>$ RtoK.
Notably, this rank ordering is reproduced by Mpipi, as shown
in Fig.~6b of ref.~\citen{Mpipi} and Figs.~6 and 7 here (red curves are 
higher than green curves for Mpipi), with
$\chi^{\rm (RtoK)}_{\rm eff}/\chi^{\rm (FtoA)}_{\rm eff}=0.95$ for
Mpipi. By comparison, KH does not capture this trend entirely,
as its predicted rank ordering of LLPS propensities is 
WT $>$ CS $>$ RtoK $>$ FtoA 
($\chi^{\rm (RtoK)}_{\rm eff}/\chi^{\rm (FtoA)}_{\rm eff}=1.26$ for KH),
as shown in Fig.~4 of ref.~\citen{SumanPNAS} and in Figs.~6 and 7 here
(red curves are lower than green curves for KH).
\\


{\bf Field-Theoretic Simulations of Phase Behaviors of Model Hydrophobic-Polar 
Sequences with the Same Composition but Different Sequence Patterns.}
As a first
test of the FTS approach introduced above for modeling sequence-dependent
LLPS effects of spatially short-range interaction, we apply the formulation
to three model hydrophobic-polar sequences with the same 
$3/2$ hydrophobic/polar composition but different sequence patterns of
the hydrophobic and polar residues.
We adopt three 20mer ($N=20$) sequences with $f_{\rm T}=0.6$ from 
Statt et al.\cite{Statt2020} (hydrophobic composition is denoted as
$f_{\rm T}$ in this reference), namely their
``[HT]$_2$TH[T$_2$H]$_4$TH'', ``T$_3$H$_3$T$_3$H$_2$T$_3$H$_3$T$_3$'',
and ``T$_{12}$H$_8$'' 
sequences listed, respectively, as the last sequence in Fig.~S6,
the sequence in Fig.~5, and the ``micelles'' sequence in Fig.~6 
in ref.~\citen{Statt2020}. 
As mentioned above, since we use leucine and serine, respectively, 
for the hydrophobic and polar residues in the present study, we refer 
to these three sequences as LS1, LS2, and LS3 (Fig.~8a).
For computational efficiency,
we opted to use these short LS sequences without electrostatic interactions
instead of the Ddx4 IDR sequences that are more than ten times longer
for the first test of our FTS formulation.
FTS of longer sequences with more complex interactions
entails technical issues with regard to equilibration
that remains to be tackled. 

The hydrophobic and polar residues are more evenly distributed
in LS1 and LS2 than in the diblock arrangement of LS3.
Statt et al. essentially use a standard hydrophobic-polar (HP) 
potential\cite{lau1989} 
whereby only hydrophobic-hydrophobic interactions are favorable
(governed by a LJ potential in the form of
Eq.~\eqref{eq:LJ} with negative contact energy and excluded volume), 
polar-polar and polar-hydrophobic interactions are repulsive 
because of excluded volume (governed by a WCA potential\cite{WCA}).
Within this HP interaction scheme, the LLPS critical temperatures 
($T_{\rm cr}=T_c$ in their notation)
for the first and second sequences are, in model units, 
$T_{\rm cr}=0.808$ and $T_{\rm cr}=1.083$, whereas 
the third sequence apparently does not
have a critical point (Table~S1 of ref.~\citen{Statt2020}).
Since we will be applying the KH interaction scheme to the LS sequences, with 
${\cal E}_{r_\alpha,r_\beta}=-0.771,0$, and $-0.304$, respectively, for
$(r_\alpha,r_\beta)=$ (leucine, leucine), (serine, serine), (leucine, serine),
instead of the HP potential, LLPS behaviors of our LS
sequences are expected to exhibit a similar but not identical trend
to their corresponding HP-like sequences.
The present choice of using the KH potential for leucine and serine---with 
nonzero interactions for all three possible interaction pairs instead
of the HP interaction scheme which is diagonalized by construction with
only one nonzero interaction---is motivated
by our desire to test the new FTS formulation
with a more complex interaction scheme that requires diagonalization
of the $\varepsilon_{r,r'}$ matrix.

\begin{figure}[!ht]
\centering
   \includegraphics[width=0.65\columnwidth]{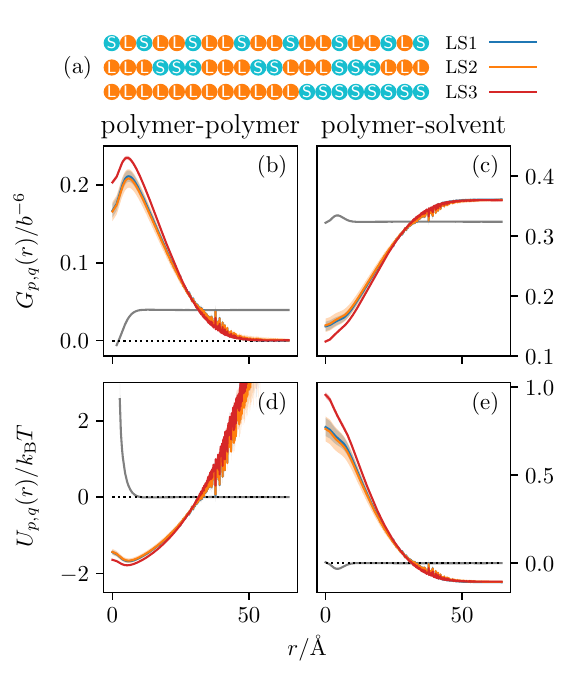}
\vskip -2 mm
\caption*{\footnotesize {\bf Fig.~8:} 
Sequence-specific LLPS effects of hydrophobic-like short-spatial-range 
interactions.
(a) Leucine-Serine (LS) 20mer polymer sequences used for FTS in this work.
(b,c) Polymer-polymer and polymer-solvent pair correlation functions $G_{\rm
b,b}(r)$ and $G_{\rm b,s}(r)$, defined in Eq.~\eqref{eq:corr_func_def}, computed
in FTS ($p={\rm b}$, $q={\rm s}$). 
(d,e) PMFs computed from the pair-correlation functions in (a,b) using
Eq.~\eqref{eq:pmf_def}. Solid curves and shaded bands in panels (b--e) depict
the average and standard deviation of twenty one 
independent simulations. The gray
solid curve in each panel corresponds to $l_{\rm h} = 0.03$ {\AA}, wherein
short-spatial-range interactions are not sufficiently strong 
to drive condensate formation
and for which no visible sequence dependence is observed. Colored curves are
computed at $l_{\rm h}=0.5$ {\AA} for which the strong short-spatial-range 
interactions cause the polymers to form a dense condensate.
}
\label{ref:fig8}
\end{figure}

The FTS is performed on an $M=48$ lattice with lattice spacing
$\Delta x = a_{\rm s} = b / \sqrt{6} \approx 1.6$ {\AA}, which amounts to a
simulation box with a linear dimension of 
$\approx 75$ {\AA}. Using the KH interaction matrix,
we truncate the summations over eigenvalues at $\tilde{a}=3$
[Eqs.~\eqref{eq:eps_spectral_decomposition} and \eqref{eq:truncated}]. 
The range of spatially short-range interactions governed by $V_{\rm h}(r)$ in 
Eq.~\eqref{eq:yukawa} is set to $\kappa^{-1}=5$ {\AA} and we let 
$v_{\rm b}=v_{\rm w}=1$ for simplicity. 
All simulations are performed without salt, and since the LS sequences
carry no electric charge, the field $\psi(\rr)$
in Eq.~\eqref{eq:field_hamiltonian} can be omitted. 
The total density of the system is set to
$\rho_0 = \rho_{\rm w} + N \rho_{\rm p} = 55.5$ M, with a polymeric
(LS sequence) component of $\rho_{\rm p} = 300$ mM giving a polymer 
bead volume fraction of $\phi=0.11$. The system thus 
contains $n_{\rm p} \approx 75$ LS polymer chains and
$n_{\rm w} \approx 1.2\times 10^4$ solvent molecules. We set the compressibility
to $\gamma=0.1 b^{-3} \approx 3.0$ M 
because of equilibration issues observed in FTS
of exactly incompressible systems.\cite{wessen2021} However, we note that
incompressible systems can alternatively be explored in FTS using the partial
saddle-point approximation of ref.~\citen{Matsen2021}. 
Simulations are performed
at $l_{\rm h} = 0.5$ {\AA} ($\chi_{\rm eff}=2.16$) and at $l_{\rm h} =
0.03$ {\AA} ($\chi_{\rm eff}=0.129$) corresponding, respectively, to 
strong and weak short-spatial-range interactions. The CL evolution equations in
Eq.~\eqref{eq:CL_evolution} are numerically solved with a time-step $\Delta t
= 5 \times 10^{-4} b^3$ using a semi-implicit integration 
scheme\cite{Lennon2008} generalized to multiple fields.\cite{mimb2022} 
We perform
21 independent simulations for each sequence and parameter set, and use
$Q_{\rm p,w}$ to monitor system equilibration. At $l_{\rm h}= 0.5$ {\AA},
the system requires $4\times 10^5$ time-steps for equilibration, after which the
field configuration are sampled every 50th step during the production run
consisting of another $4 \times 10^5$ CL steps. At $l_{\rm h}=0.03$ {\AA},
$2\times 10^4$ integration steps are sufficient for equilibration, followed by
sampling every 50th step during a production run of $3 \times 10^5$ steps. 

Fig.~8b--e shows pair-correlation functions and PMFs 
computed in FTS using Eqs.~\eqref{eq:corr_func_def} and \eqref{eq:pmf_def}
for the three LS sequences.
The grey curve in each panel corresponds to 
$l_{\rm h}=0.03$ {\AA}, 
for which the system remains in a homogeneous (single-phase) state
exhibiting no significant sequence dependence in the properties
simulated.
This single-phase state is characterized by all PMFs approaching zero at large
$r$. Colored curves in Fig.~8 correspond to $l_{\rm h}=0.5$ {\AA}, at which
the short-spatial-range interactions are sufficiently strong to drive 
the system into an inhomogeneous state containing a single spherical 
polymer-dense condensate against a near polymer-empty background. 
This feature is characterized
by the $G_{\rm p,p}(r) \sim 0$ (Fig.~8b) at separations $r$ larger than the
condensate diameter, combined with polymer-solvent anti-correlation
at small $r$ because solvent are being pushed out of the condensate.

Sequence-dependent effects are exhibited by the correlation functions
and the PMFs.
While the sequences LS1 and LS2 behave very similarly (blue and
orange curves essentially overlap in Fig.~8b--e), the
diblock sequence LS3 stands out clearly by forming denser condensates 
than both LS1 and LS2 (at small $r$, the red curve is higher than the 
orange and blue curves for correlation functions in Fig.~8b; and 
the red curve is lower than the orange and blue curves for PMFs in Fig.~8d). 
We also find that the LS3 condensate possesses a more 
heterogeneous internal structure with distinct sub-regions populated 
predominantly by either L or S residues. This
feature can be seen in the simulation snapshots in Figs.~9 and 10, which
depict the real component of the field operators
\begin{equation}
\tilde{\rho}_X(\rr) = \frac{\rho_{\rm p}}{Q_{\rm p}} \sum_{\alpha = 1}^N
\e^{W_{\alpha}(\rr)} q_{\rm F}(\rr,\alpha) q_{\rm B}(\rr,\alpha)
\delta_{r_{\alpha},X} , \quad X = \mbox{L (leucine)}, \mbox{S (serine)} ,
\end{equation}
for the number density of L and S residues
(the Kronecker $\delta_{r,r'}=1$ for $r=r'$ and 
$\delta_{r,r'}=0$ for $r\neq r'$).
To facilitate visualization of the condensate structure,
the snapshots in Fig.~9 are averaged over $10^3$ 
consecutive CL time steps.
Fig.~10 provides cross-sectional views through the center of mass
of each of the snapshots in Fig.~9. Leucine and serine densities
are represented by contours in Fig.~10 following the style introduced
by Fig.~3 of ref.~\citen{Pal2021}.
The contour plots in Fig.~10 demonstrate a marked increased degree
of structural heterogeneity of the LS3 condensate relative to the LS1 and
LS2 condensates. For LS3, the serine (polar) residues are predominantly
on the outside whereas the leucine (hydrophobic) residues are predominantly
inside. This phenomenon is reminiscent of micellar structure and, in
this regard, consistent with the simulated structure of the corresponding 
hydrophobic-polar sequence in ref.~\citen{Statt2020}.
A similar behavior has been observed in Monte-Carlo simulations of 10mer
hydrophobic-polar sequences, where the sequence of alternating 
hydrophobic-polar residues underwent LLPS whereas
the diblock condensate exhibited micellar structure instead.\cite{irback2020}
\\

$\null$

\begin{figure}[!t]
\centering
   \includegraphics[width=0.75\columnwidth]{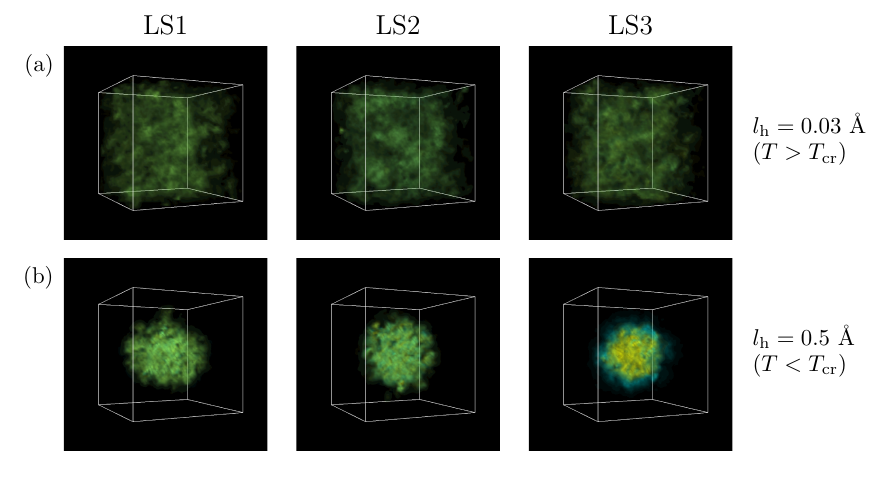}
\vskip -2 mm
\caption*{\footnotesize {\bf Fig.~9:} 
Field snapshots illustrate LLPS of LS sequences. 
Snapshots of FTS after equilibration, showing field
operators corresponding to leucine and serine number densities in orange and
cyan, respectively. The left, middle and right columns are for the sequences
LS1, LS2 and LS3, respectively, in Fig.~8a. (a) Top panels show
snapshots under conditions where interactions are not sufficiently
strong to cause condensation ($T>T_{\rm cr}$, where $T_{\rm cr}$ is 
upper critical solution temperature, UCST). (b) Bottom panels depict 
systems under conditions where there are sufficiently strong 
interactions for phase separation ($T<T_{\rm cr}$). 
All snapshots are averaged over $10^3$ CL steps. 
}
\label{ref:fig9}
\end{figure}

\begin{figure}[!ht]
\centering
   \includegraphics[width=0.75\columnwidth]{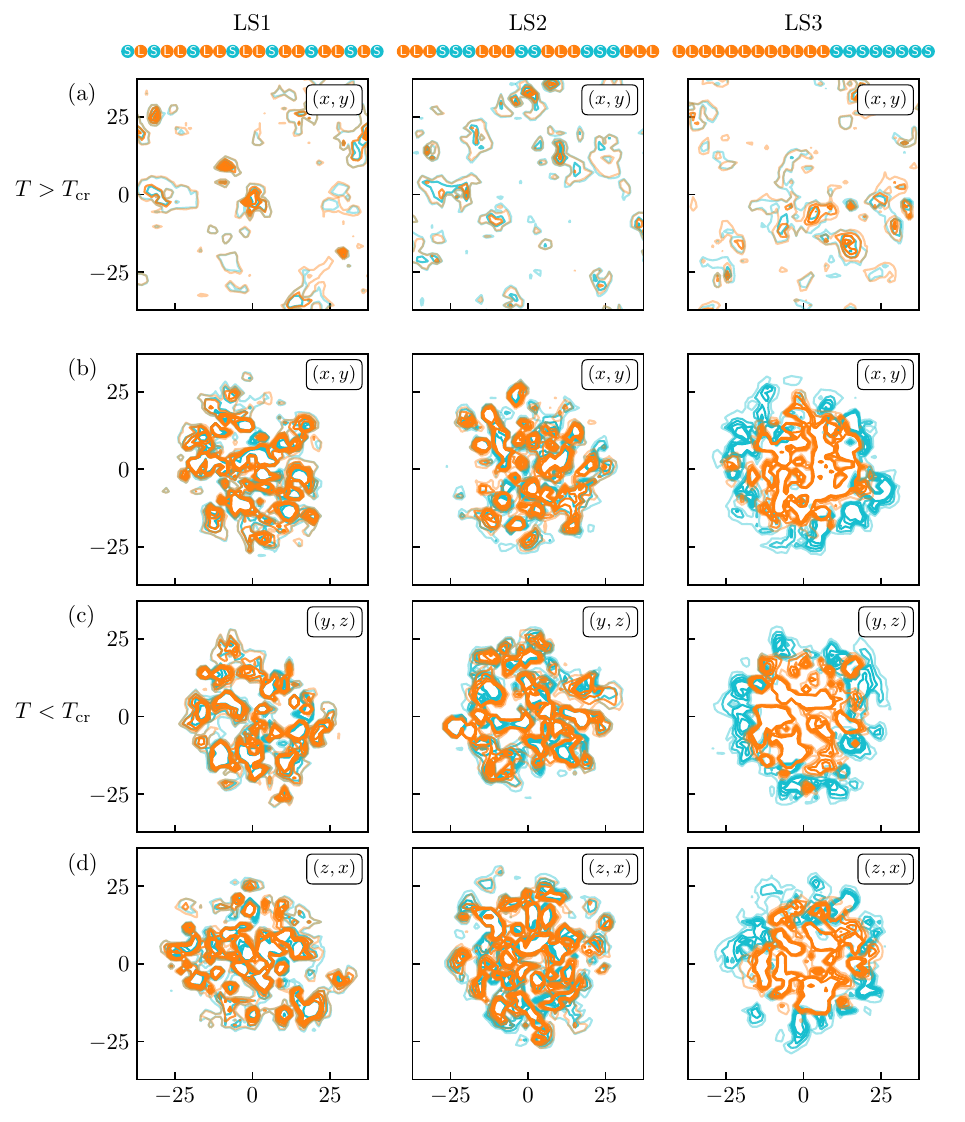}
\vskip -2 mm
\caption*{\footnotesize {\bf Fig.~10:} 
Internal structure of condensates of model hydrophobic-polar sequences.
Cross-sectional density distributions of the
snapshots in Fig.~9 in planes indicated by the insets, with $x$, $y$, and $z$
coordinates plotted in units of \AA. Leucine and serine
densities are shown in orange and cyan, respectively, as in Fig.~9. 
(a) Snapshots at $l_{\rm h} = 0.03$ {\AA} where the systems 
remain in a single-phase state. (b--d)
Snapshots at $l_{\rm h} = 0.5$ {\AA} where the LS polymers coalesce into a
near-spherical condensate.  
}
\label{ref:fig10}
\end{figure}

{\bf Assessing Analytical and FTS Results by Coarse-Grained Explicit-Chain MD.}
We now proceed to further evaluate the effectiveness of the above 
analytical and FTS approaches by comparing their predictions with the 
corresponding results from coarse-grained explicit-chain models.
As described in Models and Methods, while field-theoretic
formulation and coarse-grained modeling aim to capture essential
physics of the same system, details of the two types of models differ.
For instance, on one hand, the functional form of residue-residue interactions
is practically limited in field-theoretic formulation because of 
mathematical consideration---in the present case we are restricted 
to a single Yukawa form for all residue pairs, i.e., using $V_{\rm h}(r)$ in
Eq.~\eqref{eq:yukawa} with the same $l_{\rm h}$ and $\kappa$ for
all amino acid pairs. 
In contrast, there is more freedom in selecting
functional forms for potential energy (e.g., Lennard-Jones,
Ashbaugh-Hatch, Wang-Frenkel) 
as well as using different bead sizes to represent different amino 
acid residues; e.g., different $\sigma_{\alpha,\beta}$ 
for different types of amino acid pairs $r_\alpha,r_\beta$
in Eq.~\eqref{eq:UhhMpipi}. 
On the other hand, it is computationally much more efficient to account
for explicit solvent in field-theoretic approaches than in MD.
This recognition notwithstanding,
to be useful as a computationally efficient approach 
complementary to coarse-grained explicit-chain MD,
field-theoretic method applied to any given system should produce results 
that are in agreement, at least semi-quantitatively, with those obtained
from an appropriately constructed coarse-grained explicit-chain model
for the same system.
In future investigations, it will be interesting to explore possible
tuning of energetic parameters in our analytical and FTS formulations
to optimize agreement with coarse-grained explicit-chain MD and/or
experimental data to enable more broadly practical applications 
of field-theoretic techniques.
\\

{\bf Coarse-Grained Explicit-Chain MD for Ddx4 IDRs.}
With this in mind, phase diagrams for the four Ddx4 IDRs are obtained
by coarse-grained explicit-chain MD using the Mpipi,\cite{Mpipi} 
Urry,\cite{Urry-Mittal} and FB\cite{FB} interaction schemes (Fig.~11). 
Corresponding coarse-grained explicit-chain MD phase diagrams
using KH and HPS have been provided, respectively, in 
Fig.~4 and Fig.~3B of ref.~\citen{SumanPNAS}. The trends exhibited by
these coarse-grained MD results are largely consistent
with the analytical MFT/RPA results in Figs.~6 and 7, suggesting
that the MFT/RPA formulation is a computationally efficient approach
for exploring sequence-dependent LLPS of IDPs.
In particular, in agreement with Fig.~6b of the original Mpipi
study\cite{Mpipi}
(dotted curves in Fig.~11a, which were computed by 
using $\kappa_{\rm D}^{-1}=7.95$ {\AA}, $\epsilon_{\rm r}=80$
instead of the $\kappa_{\rm D}^{-1}=10$ {\AA}, $\epsilon_{\rm r}=40$
values used for the solid curves in Fig.~11a here)
and consistent with the present analytical MFT/RPA results in 
Fig.~7 at lower model temperatures $T\lesssim 250$~K, 
Fig.~11a exhibits the experimentally 
correct WT $>$ CS $>$ FtoA $>$ RtoK
rank ordering of LLPS propensities,\cite{Nott15,jacob2017,robert}
though the Mpipi-computed difference in LLPS propensity between 
WT and CS ($\sim 6$--$10$ K difference in $T_{\rm cr}$)\cite{Mpipi} 
is significantly smaller than the experimental estimation 
of $\sim 80$ K difference
in $T_{\rm cr}$ at [NaCl] = 100 mM (ref.~\citen{jacob2017}).
For Urry, the LLPS propensity rank ordering
WT $>$ FtoA $\approx$ RtoK $>$ CS
computed by coarse-grained MD (Fig.~11b)
is consistent with that for Urry in MFT/RPA Figs.~6 and 7.
Similarly, for FB, the RtoK $>$ WT $>$ FtoA $>$ CS
rank ordering in Fig.~11c
is consistent with that for FB in Figs.~6 and 7.
A comparison of the global critical temperatures computed by analytical
MFT/RPA and those simulated using coarse-grained MD
at a fixed monovalent salt concentration of $\sim 100$ mM
for the interaction schemes Mpipi, Urry, and FB 
is provided in Fig.~12.

\begin{figure}[!t]
\centering
   \includegraphics[width=0.85\columnwidth]{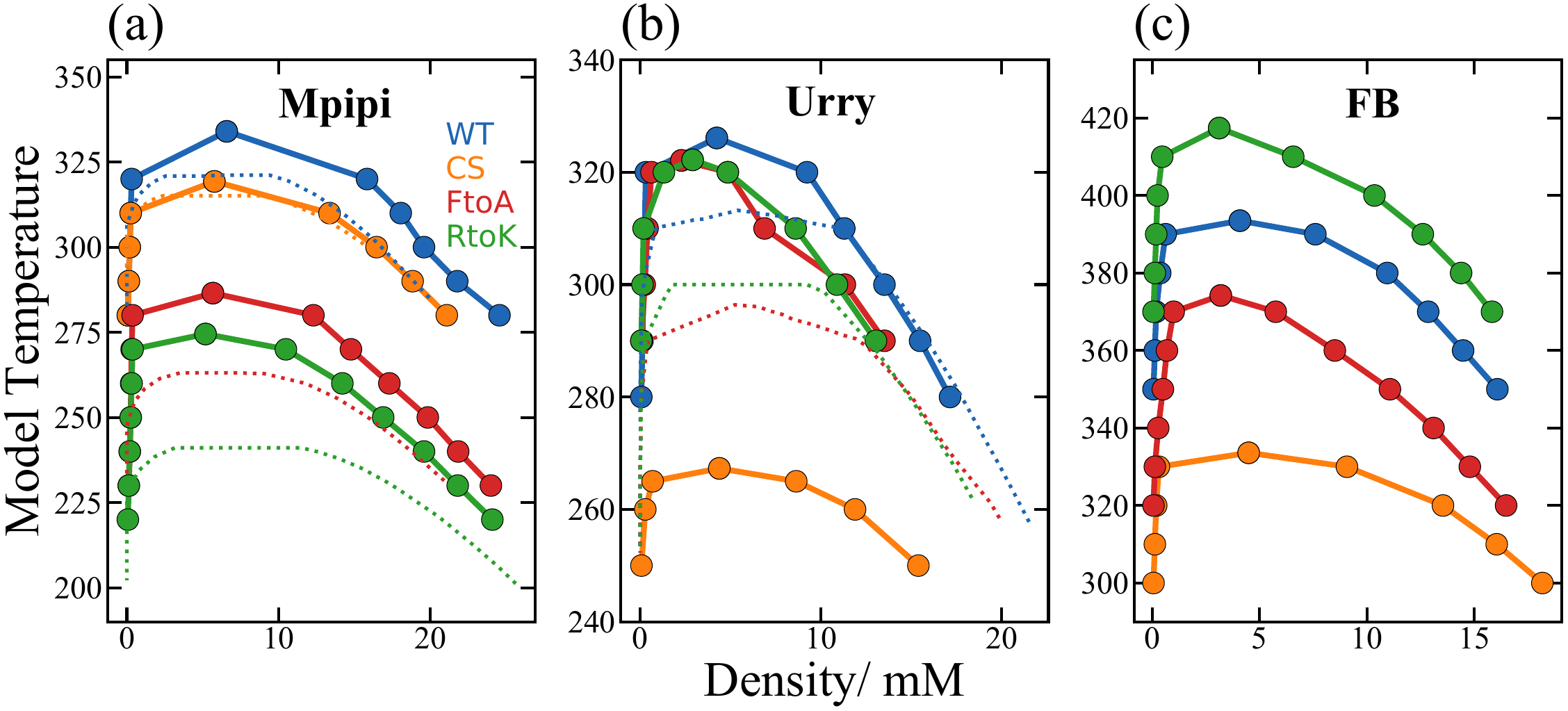}
\vskip -2 mm
\caption*{\footnotesize {\bf Fig.~11:} 
Comparing explicit-chain model phase behaviors of Ddx4 IDRs in
different interaction schemes.
Phase diagrams (coexistence curves) for WT (blue), CS (orange), FtoA (red), 
and RtoK (green) are obtained by coarse-grained MD
at $\epsilon_{\rm r}=40$ using the (a) Mpipi,\cite{Mpipi} 
(b) Urry,\cite{Urry-Mittal} or
(c) FB\cite{FB} interaction schemes in
accordance with the modeling details and simulation procedure described
in Models and Methods.
Solid lines passing through simulation data points and extrapolated
critical points\cite{dignon18,SumanPNAS,panag2017} (filled circles) 
are a guide for the eye.
Dotted lines (same color code) are previously simulated
coexistence curves adapted from
Fig.~6b of ref.~\citen{Mpipi} (a) and
Fig.~4a of ref.~\citen{Urry-Mittal} (b)
to facilitate comparison.
}
\label{ref:fig11}
\end{figure}

As discussed in previous works,\cite{SumanPNAS,Urry-Mittal,FB,Mpipi}
some of the salient differences among interaction schemes
can be understood semi-quantitatively by simple considerations
of the schemes' contact energies.
For instance, the differences in LLPS behavior of the RtoK variant 
across different interaction schemes are clearly related to
the average strength of interactions involving R versus that involving K
in the interaction schemes. For Mpipi, consider the average
${\cal E}_{\rm R,K}\equiv\sum_{r=1}^{20}{\cal E}_{r,r'}/20$ over
the 20 amino acid types, where $r'$ 
is the label for arginine (R) or lysine (K). Entries for
${\cal E}_{r,r'}$ in Supplementary Table~11 of ref.~\citen{Mpipi}
yield ${\cal E}_{\rm R}=0.2212$ kcal mol$^{-1}$ and
${\cal E}_{\rm K}=0.0563$ kcal mol$^{-1}$, thus
$\Delta{\cal E}_{\rm R-K}\equiv{\cal E}_{\rm R}-{\cal E}_{\rm K}=$
$0.1649$ kcal mol$^{-1}$ may be used to characterize the degree to
which interactions involving R are more favorable than those involving K.
The corresponding quantity in KH is
$\Delta(\lambda^{\rm hh}{\cal E})_{\rm R-K}=0.0829$ kcal mol$^{-1}$,
which is significantly smaller. Here, for KH,
$\Delta(\lambda^{\rm hh}{\cal E})_{\rm R-K}\equiv
(\lambda^{\rm hh}{\cal E})_{\rm R}-(\lambda^{\rm hh}{\cal E})_{\rm K}$,
with $\lambda^{\rm hh}{\cal E}_{\rm R,K}\equiv
-\sum_{r=1}^{20}\lambda^{\rm hh}_{r,r'}{\cal E}_{r,r'}/20$, 
and $-\lambda^{\rm hh}_{r,r'}{\cal E}_{r,r'}$ 
being the entries in Table~S3 of ref.~\citen{dignon18}.
For HPS, Urry, and FB,
the corresponding difference in interaction strength is given by
${\cal E}\Delta\lambda_{\rm R-K}/2\equiv
{\cal E}[\lambda({\rm R})-\lambda({\rm K})]/2$ where 
${\cal E}=0.2$ kcal mol$^{-1}$ and $\lambda({\rm R})$ and $\lambda({\rm K})$ 
are, respectively, the value for R and K on a given 
hydrophobicity/hydropathy scale (see Models and Methods).
From the scales in Table~S1 of ref.~\citen{dignon18},
the ``Urry et al.'' column of Table~S2 of ref.~\citen{Urry-Mittal}, 
and Table~S7 of ref.~\citen{FB},
${\cal E}\Delta\lambda_{\rm R-K}/2=$
$-0.0514$, $0.0176$, and $-0.0231$ kcal mol$^{-1}$, respectively,
for HPS, Urry, and FB.
Taken together, these numbers for Mpipi, KH, Urry, HPS, and FB, viz.,
$0.1807$, $0.0829$, $0.0176$, $-0.0459$, and $-0.0170$
respectively, provide a useful explanation
for the signs and magnitudes of their WT $-$ RtoK critical temperature 
differences $\Delta(T_{\rm cr})_{\rm WT-RtoK}
\equiv(T_{\rm cr})_{\rm WT}-(T_{\rm cr})_{\rm RtoK}$ 
$\approx$ $60$, $43$, $4$, $-29$, and $-24$ K 
as seen for Mpipi, KH, Urry, 
HPS, and FB, respectively, in Fig.~11 of the present study as well as
in Fig.~3B and Fig.~4 of ref.~\citen{SumanPNAS}.

\begin{figure}[!t]
\centering
   \includegraphics[width=0.75\columnwidth]{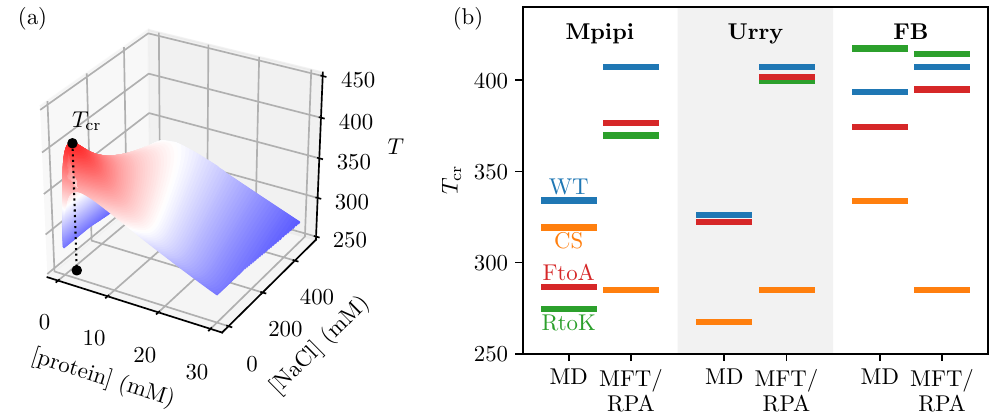}
\vskip -2 mm
\caption*{\footnotesize {\bf Fig.~12:} 
Comparing the critical temperature $T_{\rm cr}$ in the coarse-grained
explicit-chain MD and the analytical MFT/RPA models.
(a) An example of [protein]-[salt]-temperature phase diagrams computed 
in the approximate analytical MFT/RPA theory 
for WT Ddx4 IDR using Mpipi.
The depicted binodal surface is constructed by stacking
[protein]-[salt] coexistence curves at constant $T$ (i.e.~solid curves in
Fig.~6) at heights corresponding to their respective temperatures. The
indicated global critical temperature, $T_{\rm cr}$, is found by maximizing the
function $T(\rho_{\rm p},\rho_{\rm s})$ defined implicitly through $\det
\mathcal{H} = 0$ where $\mathcal{H}$ is a $2\times 2$ Hessian matrix 
with entries
$\partial^2 f / \partial\rho_i \partial\rho_j$, and $i,j={\rm p,s}$, 
(ref.~\citen{mimb2022}). 
The vertical dotted line shows the projection of the critical point onto the
[protein]-[salt] plane, indicating the corresponding critical densities. (b)
The rank orderings of $T_{\rm cr}$ for the Ddx4 IDRs
in coarse-grained MD in Fig.~11 
and the global $T_{\rm cr}$ in MFT/RPA 
mostly agree for each of the three interaction schemes. Note that
the discrepancies in the magnitudes of MD- and MFT/RPA-predicted $T_{\rm cr}$
values shown here are largely
a consequence of different energy scales used for MD
and MFT/RPA. As stated in the discussion above for the results in Fig.~6,
in order to facilitate comparison across different interaction schemes,
the chosen energy scales for MFT/RPA are such that 
$\chi_{\rm eff}$ is always equal to $0.5$ at a reference temperature $T=300$K 
and $l_{\rm B} = 7$ \AA~ for WT Ddx4 for all interaction schemes (thus all
the blue lines for MFT/RPA in (b) are at the same level).
In contrast, we largely follow the full interaction potentials provided 
in refs.~\citen{Urry-Mittal,FB} and \citen{Mpipi}
for the Urry, FB, and Mpipi interaction schemes. 
It follows that only the rank orderings---but not the magnitudes---of
$T_{\rm cr}$s for the Ddx4 variants are comparable in (b).
}
\label{ref:fig12}
\end{figure}

As noted above, the difference in LLPS propensity between
WT and CS predicted by Mpipi is small, with 
$\Delta(T_{\rm cr})_{\rm WT-CS}\approx 6$ K 
in the original simulation\cite{Mpipi} and
$\approx 15$ K in Fig.~11a, both much less than the
experimentally estimated 
$\Delta(T_{\rm cr})_{\rm WT-CS}
\sim 80$ K (refs.~\citen{jacob2017}).
In fact, the slightly higher 
$\Delta(T_{\rm cr})_{\rm WT-CS}$ we obtain
is probably due in large part to the smaller dielectric constant
$\epsilon_{\rm r}=40$ used for Fig.~11a instead of the $\epsilon_{\rm r}=80$
value used in the original study.
By comparison, larger $\Delta(T_{\rm cr})_{\rm WT-CS}$ values are predicted by 
most of the other interaction schemes considered here:
For HPS, $\Delta(T_{\rm cr})_{\rm WT-CS}\approx 14$ K 
for $\epsilon_{\rm r}=80$ (Fig.~3B of ref.~\citen{SumanPNAS}).
For KH, $\Delta(T_{\rm cr})_{\rm WT-CS}\approx$ 21 and 46 K, respectively,
for $\epsilon_{\rm r}=80$ and $40$ (Fig.~4 of ref.~\citen{SumanPNAS}).
For Urry and FB simulated here at $\epsilon_{\rm r}=40$, 
$\Delta(T_{\rm cr})_{\rm WT-CS}\approx 60$ K (Fig.~11b,c).
Presumably, this conspicuous difference between Mpipi and the other interaction 
schemes regarding CS properties is chiefly caused by Mpipi's electric
charge assignment of $0.75e$ for R and K, $-0.75e$ for aspartic acid (D) 
and glutamic acid (E), and $0.375e$ for histidine\cite{Mpipi} instead of using
the full proton charge $e$ for electrostatic interactions among
R, K, D, and E as in
the other interaction schemes. Since each of the Ddx4 IDRs
contains only two histidines, the main effects are in R, K, D, and E.
The assignment of $\pm 0.75e$ electric charges instead of $\pm e$
amounts to a reduction factor $(0.75)^2=0.5625$
in electrostatic interaction strength, which is further attenuated by
Mpipi's adoption of a slightly shorter Debye screening length 
$\kappa_{\rm D}^{-1}=7.95$ {\AA} instead of the $\kappa_{\rm D}^{-1}=10$ {\AA} 
employed for the other interaction schemes. Because the difference
in LLPS propensity between WT and CS is principally a sequence-specific
electrostatic effect, Mpipi's reduced electrostatic strength
should go a long way in accounting for the relatively small
$\Delta(T_{\rm cr})_{\rm WT-CS}$ it predicts.

Secondary to electrostatics, another possible origin of 
the observed variation in $\Delta(T_{\rm cr})_{\rm WT-CS}$ 
predicted by different interaction schemes is their
different spatially short-range nonelectrostatic residue-residue
contact energies. WT is transformed into CS, and vice versa, by swapping 
residues among 24 sequence positions---involving 6 D, 6 E, 9 R, and 3 K
residues participating in 4 (E $\leftrightarrow$ R), 5 (D $\leftrightarrow$ R),
2 (E $\leftrightarrow$ K), and 1 (D $\leftrightarrow$ K) interchanges.
Because the D, E, R, and K residues engage in nonelectrostatic as well as
electrostatic interactions, 
sequence-specific nonelectrostatic interactions do contribute
to $\Delta(T_{\rm cr})_{\rm WT-CS}$ and this
effect clearly depends on the interaction scheme.
Now, the energetic effects of the (E $\leftrightarrow$ R), 
(D $\leftrightarrow$ R), (E $\leftrightarrow$ K), and (D $\leftrightarrow$ K) 
swaps may be characterized by 
${\cal E}\Delta\lambda_{\rm E-R}/2$, ${\cal E}\Delta\lambda_{\rm D-R}/2$,
${\cal E}\Delta\lambda_{\rm E-K}/2$, and ${\cal E}\Delta\lambda_{\rm D-K}/2$;
but these quantities can have different magnitudes and even different
signs for different interaction schemes, as exemplified by the
${\cal E}\Delta\lambda_{\rm E-R}/2$, ${\cal E}\Delta\lambda_{\rm D-R}/2$,
${\cal E}\Delta\lambda_{\rm E-K}/2$, and ${\cal E}\Delta\lambda_{\rm D-K}/2$
values of
$0.0459$, $0.0378$, $-0.0055$, and $-0.0136$ kcal mol$^{-1}$, respectively,
for HPS, the corresponding values of
$-0.0558824$, $-0.0264705$, $-0.0382354$, and $-0.0088235$ kcal mol$^{-1}$ 
for Urry, and $0.018596$, $0.0065$, $-0.004485$, and $-0.016581$ for FB.
Ramifications of these differences on the LLPS properties of model CS Ddx4
IDR remain to be further explored.
\\

{\bf Coarse-Grained Explicit-Chain MD for Hydrophobic-Polar Sequences.}
Fig.~13 provides phase behaviors computed by explicit-chain MD simulation
for the three LS sequences studied above by FTS.
The rank ordering of LLPS propensity LS3 $>$ LS2 $\gtrsim$ LS1 observed
in Fig.~13a is consistent with that deduced from FTS correlation 
functions and PMFs in Fig.~8, attesting to the effectiveness
of the FTS formulation developed here for sequence-specific effects
of spatially short-range interactions on LLPS of IDPs.
There are minor differences between the FTS and MD results, in that
the FTS correlation functions and PMFs suggest that the LLPS
propensities of LS1 and LS2 are practically identical considering
numerical uncertainties, but the MD phase diagrams show that LS2 has a slightly
higher LLPS propensity than LS1. Such minor differences are not
unexpected given the differences discussed above between 
the FTS and MD models. In both our FTS and MD models, the more blocky
LS3 sequence clearly exhibits its superior LLPS propensity.

\begin{figure}[!t]
\centering
   \includegraphics[width=0.85\columnwidth]{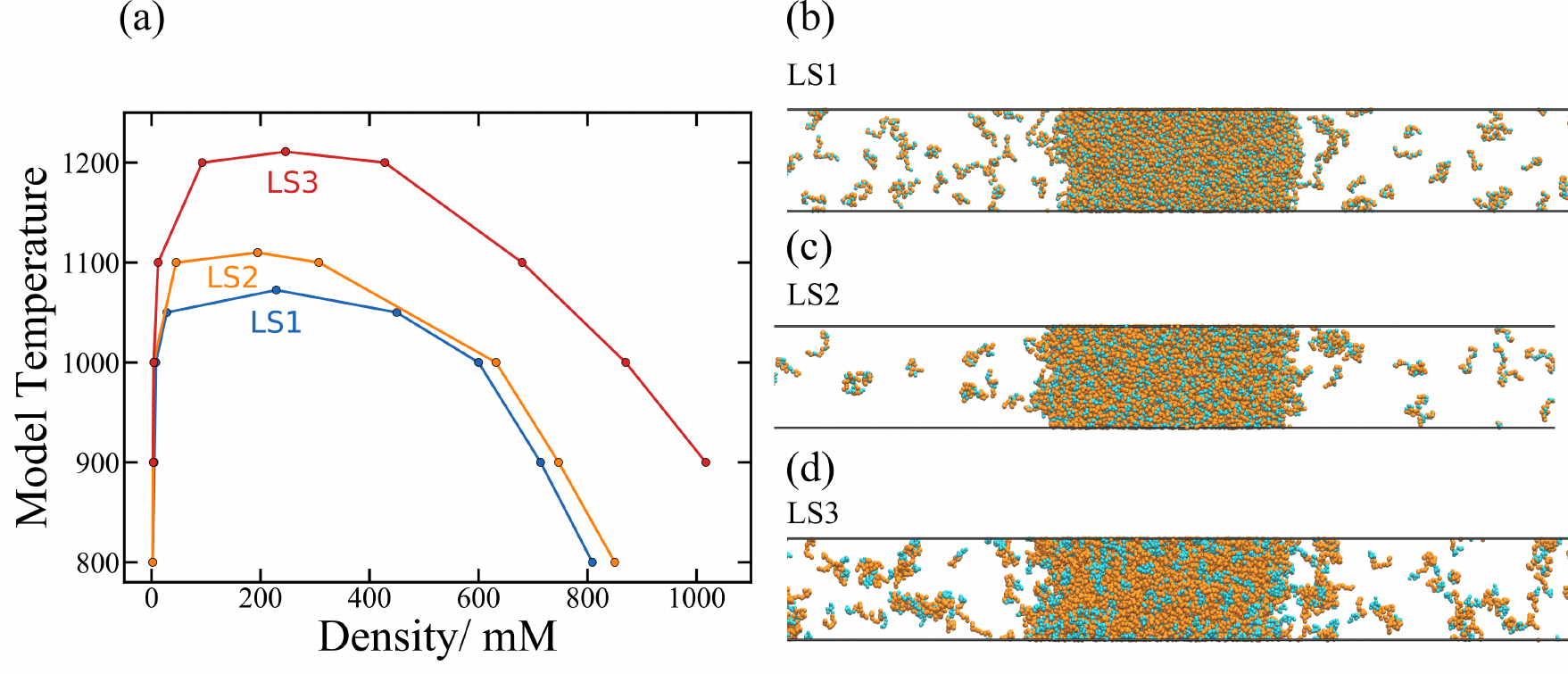}
\vskip -2 mm
\caption*{\footnotesize {\bf Fig.~13:} 
Sequence-dependent phase behaviors of hydrophobic-polar sequences 
in our explicit-chain model.
(a) Phase diagrams (coexistence curves) of the three leucine-serine sequences
LS1, LS2, and LS3 in Fig.~8a are computed by coarse-grained
MD using the KH interaction scheme\cite{dignon18} and the protocol described in 
Models and Methods.
Simulation data points and extrapolated critical 
points\cite{dignon18,SumanPNAS,panag2017} are shown as filled circles, 
connecting lines are a guide for the eye. 
(b--d) Snapshots taken during the simulation showing a part of the 
simulation box for 1,000 copies of
(b) LS1 at model temperature $T=1000$ K, (c) LS2 at model $T=1000$ K, and
(d) LS3 at model $T=1200$ K. 
Beads representing leucine and serine are shown, respectively, in orange 
and cyan in these snapshots. 
Significant number of polymers are seen in the dilute phase of each 
of these systems.
}
\label{ref:fig13}
\end{figure}

The critical temperatures of the LS sequences shown in Fig.~13a are 
$3$--$4$ times higher than those of the Ddx4 IDRs in Fig.~11, exhibiting 
stable droplets at model temperatures as high as $1000$ K (Fig.~13b--d). 
This observation is notable because LLPS propensities tend to increase 
with chain length\cite{linPRL} but the LS sequences have a much shorter 
chain length ($N=20$) than that of the Ddx4 IDRs ($N=236$).
A likely reason for the high LLPS propensities of the LS sequences
is the strongly favorable leucine-leucine interaction
(strongest in the KH scheme) and favorable leucine-serine interactions 
in the model. The snapshot in Fig.~13d for LS3 shows a certain degree of
separate clustering of the leucine residues and of the serine residues,
but not to the extent displayed by the micellar organization of
the corresponding hydrophobic-polar sequence in ref.~\citen{Statt2020},
probably because the leucine-serine interaction is favorable in our KH 
interaction scheme (and thus conducive to leucine-serine mixing) but the 
interaction between a hydrophobic and a polar bead is repulsive 
in ref.~\citen{Statt2020}.

A sequence hydropathy decoration (SHD) parameter, 
defined as SHD $=N^{-1}\sum_{\alpha<\beta}^N
[\lambda(r_\alpha)+\lambda(r_\beta)]|\alpha-\beta|^{-1}$,
was proposed recently as a predictor of properties of single-chain 
conformational ensembles of heteropolymers.\cite{zhengHP}
For applications to interaction schemes such as KH and Mpipi that have
210 contact energies instead of a 20-value hydrophobicity/hydropathy
scale, we consider a natural generalization of the above formula:
\begin{equation}
\label{eq:SHDhh} 
{\rm SHD} \rightarrow
{\rm SHD}^{\rm hh} \equiv
-N^{-1}\sum^N_{\alpha<\beta}\varepsilon_{r_\alpha,r_\beta}|\alpha-\beta|^{-1}
\; ,
\end{equation}
where $\varepsilon_{r,r'}$ for KH is given by the entries in
Table~S3 of ref.~\citen{dignon18}. 
For the LS1, LS2, and LS3 sequences we consider here in the KH interaction 
scheme, ${\rm SHD}^{\rm hh}=1.12$, $1.05$, and $1.17$, respectively.
As the ${\rm SHD}^{\rm hh}$ value of LS3 with higher LLPS propensity
is larger than those of LS1 and LS2, this result suggests that 
SHD or ${\rm SHD}^{\rm hh}$ may be used as a predictor for LLPS
as well. Further effort will be needed to examine whether the
correlation between SHD or ${\rm SHD}^{\rm hh}$ and LLPS propensity of 
heteropolymers with spatially short-range interactions is as strong as
that between sequence charge decoration (SCD)\cite{kings2015}
and LLPS propensity of polyampholytes.\cite{lin2017,suman1,suman2}
\\

$\null$

\noindent{\large\bf CONCLUSIONS}\\

In summary, we have developed a field-theoretic formulation
for modeling sequence-specific biomolecular phase separation.
Our theory offers a coarse-grained account of short-spatial-range
$\pi$-related and hydrophobic interactions as well as long-spatial-range 
Coulomb interactions. In conjunction with RPA for electrostatics, a
mean-field approximation for spatially short-range interactions
derived from the general theory is useful as a computationally efficient 
tool for studying phase separation of intrinsically disordered proteins,
as exemplified by the application to the Ddx4 IDRs described here.
Full sequence effects of spatially short-range interactions
can be studied using FTS, illustrated here by the sequence-dependent
phase properties of three different hydrophobic-polar sequences
with the same hydrophobic/polar composition. Initial success
in applying these approaches to several different interaction schemes 
for modeling biomolecular LLPS are confirmed by coarse-grained 
explicit-chain molecular dynamics simulations. We have included 
only Yukawa potentials in our theory for this initial effort.
Because ostensibly small changes in coarse-grained interaction
potentials can lead to fundamental variations in conformational
properties,\cite{cheung2002,liu2005,kaya2013}
future effort should aim to extend our formulation to other functional 
forms for spatially short-range potentials,\cite{ottinger2021}
including temperature-dependent effects.\cite{Mittal-ACSCent2019,liu2005}
Techniques should also be further developed to study longer sequences
of biological IDRs by FTS. Much exciting work lies ahead under the
present theoretical framework.
\\


{\bf Acknowledgements.}
We thank Yi-Hsuan Lin for helpful discussions. Financial support
for this work was provided by Canadian Institutes of Health
Research grant NJT-155930 and Natural Sciences and Engineering
Research Council of Canada Discovery grant RGPIN-2018-04351 to H.S.C. 
We are grateful for the computational resources provided generously
to our research group by 
Compute/Calcul Canada and the Digital Research Alliance of Canada.
\\

\noindent The authors declare no conflict of interest.

\vskip 4cm

\vfill\eject 

$\null$
\begin{center}
   \includegraphics[height=44.5mm]{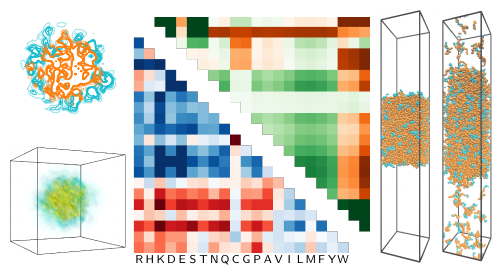}
\end{center}
\centerline{\bf TOC graphics}

\vfill\eject

\end{document}